\PassOptionsToPackage{table,xcdraw}{xcolor}
\documentclass[11pt,a4paper]{article}
\pdfoutput=1
\usepackage{jheppub}
\usepackage{gensymb}
\DeclareSymbolFont{matha}{OML}{txmi}{m}{it}
\DeclareMathSymbol{\varv}{\mathord}{matha}{118}
\usepackage{amsmath}
\usepackage{amssymb}
\usepackage{graphicx}
\usepackage{subfigure}
\usepackage{pstricks}
\usepackage{bm}
\usepackage{pbox}
\usepackage{placeins}
\usepackage{booktabs}
\usepackage[T1]{fontenc}
\usepackage{footnote}
\usepackage{pdfpages}
\usepackage{hhline}
\usepackage{multirow}
\usepackage{multicol}
\usepackage{enumitem}
\usepackage[toc,page]{appendix}
\allowdisplaybreaks
\usepackage{comment}
\usepackage{float}
\usepackage{slashed}
\usepackage{xcolor}
\usepackage{textcomp}
\usepackage{multirow}
\usepackage[numbers]{natbib}
\usepackage{notoccite}
\usepackage{soul}

\usepackage{extarrows}

\usepackage{physics}
\usepackage{comment}
\usepackage{dsfont}
\usepackage{csquotes}
\usepackage{soul}
\usepackage[none]{hyphenat}
\DeclareUnicodeCharacter{0304}{ }
\usepackage{rotating}
\usepackage{tabularx}
\usepackage{ragged2e}

\usepackage[many]{tcolorbox}

\renewcommand{\thetable}{\Roman{table}}

\definecolor{MyDarkBlue}{rgb}{0.1, 0.1, 0.8} 
\definecolor{MyLightBlue}{rgb}{0.22,0.51,0.9}
\definecolor{MyGreen}{rgb}{0.0, 0.5, 0.0}
\definecolor{BrickRed}{rgb}{0.8, 0.25, 0.33}
\definecolor{magenta}{rgb}{1.0,0.0,1.0}
\definecolor{brightgreen}{rgb}{0.4,1,0}
\definecolor{orange}{rgb}{1.0, 0.65, 0.0}

\usepackage{tikz}

\newif\ifstartedinmathmode
\newcommand\encircled[1]{%
  \relax\ifmmode\startedinmathmodetrue\else\startedinmathmodefalse\fi%
  \tikz[baseline,anchor=base]{%
  \node[color=red,draw=MyDarkBlue,circle,thick,outer sep=0pt,inner sep=.2ex]
    {\ifstartedinmathmode$#1$\else#1\fi};}%
}
\newif\ifstartedinmathmode
\newcommand\twincircled[1]{%
  \relax\ifmmode\startedinmathmodetrue\else\startedinmathmodefalse\fi%
  \tikz[baseline,anchor=base]{%
    \draw [MyDarkBlue, thick] circle [radius=.32](0,0) ;
  \node[color=red,draw=brightgreen,circle,thick,outer sep=0pt,inner sep=0]
    {\ifstartedinmathmode$#1$\else#1\fi};}%
}

\RequirePackage{hyperref}
\hypersetup{colorlinks=true, citecolor=MyGreen,linkcolor=BrickRed, urlcolor=MyLightBlue}

\title{\bf Naturally Light Dirac and Pseudo-Dirac Neutrinos from Left-Right Symmetry}
\author[a]{K.S. Babu,}
\author[b,c]{Xiao-Gang He,}
\author[a]{Mingxian Su,}
\author[d]{Anil Thapa}

\affiliation[a]{Department of Physics, Oklahoma State University, Stillwater, OK 74078, USA}
\affiliation[b]{Tsung-Dao Lee Institute, and School of Physics and Astronomy,
Shanghai Jiao Tong University, Shanghai 200240, China}
\affiliation[c]{National Center for Theoretical Sciences, and Department of Physics,
National Taiwan University, Taipei 10617, Taiwan}
\affiliation[d]{Department of Physics, University of Virginia, Charlottesville, Virginia 22904-4714, USA}

\emailAdd{babu@okstate.edu}\emailAdd{hexg@sjtu.edu.cn}\emailAdd{mx.su@okstate.edu}\emailAdd{wtd8kz@virginia.edu}

\abstract{
We develop a class of left-right symmetric theories based on the gauge group $SU(3)_c \times SU(2)_L \times SU(2)_R \times U(1)$ with a generalized seesaw mechanism for generating the charged fermion masses. Neutrinos are naturally Dirac particles in this setup with their small masses arising from two-loop quantum corrections.  We evaluate these two-loop diagrams exactly and analyze the flavor structure of the lepton sector. We find excellent fits to neutrino oscillation data, independent of the right-handed gauge symmetry breaking scale. We also explore the possibility that neutrinos are pseudo-Dirac particles in this framework, with the tiny mass splittings between active and sterile neutrinos arising from Planck-induced corrections and find possible realizations. These models can be tested in the near future with precision cosmological measurements of $\Delta N_{\rm eff}$ in CMB which is predicted to be $\simeq 0.14$.  This class of models allows for a solution to the strong CP problem via parity symmetry without the need for an axion. }

\begin{document}
\maketitle


\section{Introduction}\label{sec:Intro}

One of the fundamental questions in particle physics today is whether or not global lepton number and baryon number symmetries present in the standard model (SM) are true symmetries.  This question should be addressed in the context of SM extensions that generate small neutrino masses needed to explain the observed oscillations among different neutrino flavors.  Such extensions may or may not respect lepton number as a global symmetry.  In the framework of the popular seesaw mechanism \cite{Minkowski:1977sc,Gell-Mann:1979vob,Mohapatra:1979ia, Yanagida:1980xy, Glashow:1979nm} which explains naturally the smallness of the neutrino mass, lepton number is broken by the heavy Majorana mass of the right-handed neutrino.  However, neutrino oscillation experiments cannot distinguish a Majorana neutrino from a Dirac neutrino.  The hallmark of lepton number violation in low energy experiments is the observation of neutrinoless double beta decay, which has not been seen in experiments to date  (for a review on the current status see Ref.  \cite{Dolinski:2019nrj}).  It is therefore important to explore theories where neutrinos are Dirac particles, especially if there is a natural understanding of their small masses. 

The purpose of this paper is to develop a class of left-right symmetric theories that lead naturally to light Dirac neutrinos \cite{Babu:1988yq}.  Unlike in conventional left-right symmetric theories \cite{Pati:1974yy, Mohapatra:1974gc, Mohapatra:1974hk, Senjanovic:1975rk}, in these theories the masses of all charged fermions arise from a generalized seesaw mechanism \cite{Davidson:1987mh}.  Interestingly, with the minimal fermionic content, the seesaw mechanism is not effective in the neutrino sector, and the neutrinos remain as Dirac fermions with their small masses generated via two-loop radiative corrections.  While this class of models was proposed and the neutrino masses were estimated in Ref. \cite{Babu:1988yq}, the flavor structure of the lepton sector has not been studied thus far.  Here we undertake a detailed study of the flavor structure and confront these models with neutrino oscillation data. We evaluate exactly the two-loop diagrams generating small Dirac neutrino masses and show with our numerical analysis that these models are fully consistent with neutrino oscillation phenomenology, regardless of the scale at which the right-handed gauge symmetry breaks down spontaneously.  

This class of left-right symmetric theories with naturally light Dirac neutrinos is based on the gauge group $SU(3)_c \times SU(2)_L \times SU(2)_R \times U(1)$. These theories are motivated on several grounds.  First, parity is a spontaneously broken symmetry in this framework, unlike in the SM.  The Higgs sector of the model is very simple, consisting of an $SU(2)_L$ doublet and an $SU(2)_R$ doublet. The generalized seesaw mechanism present in these theories provides some understanding of the mass hierarchies observed among quarks and leptons \cite{Davidson:1987mh,Davidson:1987tr}.  Owing to parity symmetry present in the theory, the strong CP problem can be solved without resorting to the  Peccei-Quinn symmetry and the resulting axion \cite{Babu:1989rb}. (For studies of low energy phenomenology of these models see Ref. \cite{Babu:2018vrl,Hall:2018let,Craig:2020bnv}).\footnote{For variations of these models where small Dirac neutrino masses arise via one-loop diagrams in presence of an extended  Higgs sector, see Ref. \cite{Mohapatra:1987hh}.} These theories have a natural embedding in $SU(5)_L \times SU(5)_R$ grand unification \cite{Davidson:1987mh,Lee:2016wiy}.\footnote{While the models of Ref. \cite{Davidson:1987mh,Davidson:1987tr} were termed ``universal seesaw models", the model of Ref. \cite{Babu:1988yq} which we develop further here, does not fall into this class, since there is no seesaw mechanism acting on the neutrinos.}

While the discovery of neutrinoless double beta decay would be a definitive confirmation of the Majorana nature of the neutrino \cite{Schechter:1981bd}, there are independent cosmological tests that can distinguish a Dirac neutrino from a Majorana neutrino. The effective number of light degrees of freedom relevant for early universe evolution, parameterized as  $\Delta N_{\rm eff}$, will receive an additional contribution of about 0.14 in the Dirac neutrino mass scenario presented here, which is within the sensitivity reach of forthcoming CMB measurements SPT-3G \cite{SPT-3G:2014dbx}, SO \cite{SimonsObservatory:2019qwx} and CMB-S4 \cite{Lee:2013bxd,Abazajian:2019eic} experiments, which are expected to have a sensitivity of $0.06$ or better in $\Delta N_{\rm eff}$.

In any scenario where the neutrinos are Dirac particles with naturally small masses, there is a possibility that they could be pseudo-Dirac particles at a more fundamental level \cite{Wolfenstein:1981kw,Petcov:1982ya,Valle:1983dk,Kobayashi:2000md}.  Quantum gravity corrections could generate ultra-small Majorana masses for both the left-handed and the right-handed neutrinos via higher dimensional operators suppressed by the Planck scale.  These corrections should respect all gauge symmetries present in the theory, but are not expected to respect global symmetries such as lepton number.  We explore such a  possibility in our left-right symmetric framework. We propose to gauge the $(B-L)$ symmetry present in the theory in order to control the amount of active-sterile mass splitting that these quantum gravity corrections would induce.  The spontaneous breaking of $(B-L)$ symmetry naturally would lead to an unbroken $Z_N$ subgroup of lepton number, with the integer $N$ dependent on the $(B-L)$ charge of the  Higgs scalar that breaks the symmetry.  For $N=2$, Majorana neutrino masses are permitted, while for any other value of $N$, the neutrino would remain strictly  a Dirac particle, even with the inclusion of quantum gravity corrections.  We analyze the case of $N=2$ in some detail that leads to pseudo-Dirac neutrinos, which have been subject of much discussion in the context of ultra-high energy neutrino detection at neutrino telescopes such as IceCube \cite{Beacom:2003eu,Keranen:2003xd}, as well as in supernova neutrino signals \cite{Martinez-Soler:2021unz}. 

Naturally light Dirac neutrinos can arise in other contexts. In the mirror universe scenario the gauge symmetry of the SM is duplicated in a mirror sector \cite{Lee:1956qn,Foot:1991py,Berezhiani:1995yi}. The mirror neutrino $\nu'$ from the mirror lepton doublet $L'= (\nu',\, e')$ can then be the Dirac partner of the usual neutrino $\nu$.  A Dirac mass connecting $\nu$ and $\nu'$ would arise from a dimension-five operator $(LH)(L'H')/\Lambda$ involving the SM Higgs ($H$) and the mirror Higgs ($H'$) doublets, leading to a Dirac seesaw mechanism \cite{Silagadze:1995tr}. Other models of naturally light Dirac neutrinos have been constructed assuming certain discrete symmetries (see for e.g. Ref. \cite{Farzan:2012sa}) and/or by extending the scalar sector (see for e.g. Ref. \cite{Ma:2016mwh,Saad:2019bqf,Jana:2019mgj}). Smallness of neutrino masses is explained in these models as they arise through loop-induced quantum corrections. In contrast to these models, the model developed here is motivated on other grounds -- parity as a spontaneously broken symmetry, solution to the strong CP problem via parity symmetry, origin from $SU(5)_L \times SU(5)_R$ unification, etc, and does not require an extended Higgs sector.

The rest of the paper is organized as follows. In Sec. \ref{sec:sec2} we describe the left-right symmetric model with seesaw mechanism for charged fermions. In Sec. \ref{sec:sec3} we evaluate the two-loop diagrams for Dirac neutrino masses exactly. In Sec. \ref{sec:sec4} we carry out fits to the neutrino oscillation data within the model and show its consistency.
In Sec. \ref{sec:sec5} we discuss the embedding of the model in $SU(5)_L \times SU(5)_R$ unification. In Sec. \ref{sec:sec6} we discuss cosmological tests of these models in $N_{\rm eff}$. In Sec. \ref{sec:sec7} we discuss the possibility of realizing pseudo-Dirac neutrinos with the aid of quantum gravity corrections.  Finally we conclude in Sec. \ref{sec:sec8}.

\section{Left-right Symmetric Models with Seesaw Mechanism for Charged Fermion Masses}
\label{sec:sec2}
The class of models we develop here is based on the left-right symmetric gauge group $SU(3)_c \otimes SU(2)_L \otimes SU(2)_R \otimes U(1)_X$. The fermions transform  under the gauge group in a left-right symmetric manner which enables  parity to be defined as a symmetry of the Lagrangian:
\begin{align}
Q_L\ (3,2,1,1/3) &= \begin{pmatrix}
 u_L \\
 d_L \\
\end{pmatrix} , \hspace{5mm}
Q_R\ (3,1,2,1/3) = \begin{pmatrix}
 u_R \\
 d_R \\
\end{pmatrix} , \nonumber\\[5pt]
\Psi_L\ (1,2,1,-1) &= \begin{pmatrix}
 \nu_L \\
 e_L \\
\end{pmatrix}, \hspace{5mm}
\Psi_R\ (1,1,2,-1) = \begin{pmatrix}
 \nu_R \\
 e_R \\
\end{pmatrix} .
\label{eq:fermion}
\end{align}
Under parity symmetry $Q_L \leftrightarrow Q_R,\, \Psi_L \leftrightarrow \Psi_R$ along with $W_L \leftrightarrow W_R$ where $W_{L,R}$ denote the three gauge bosons of $SU(2)_{L,R}$. 

The model also has three families of vector-like quarks and leptons that transform as singlets of $SU(2)_L \times SU(2)_R$:
\begin{equation}
    P(3,1,1,4/3),~~~N(3,1,1,-2/3),~~~E(1,1,1,-2)~.
    \label{eq:vecferm}
\end{equation}
These fermions are utilized to generate masses for quarks and leptons via a generalized seesaw mechanism. As shown in Sec. \ref{sec:sec5}, the fermion multiplets shown in Eqs. (\ref{eq:fermion})-(\ref{eq:vecferm}) neatly fit into $\{(10,1) + (\overline{5}, 1)\}$ and $\{(1,10) + (1, \overline{5})\}$ representations  of $SU(5)_L \times SU(5)_R$ unification, which are the simplest anomaly-free chiral representations of this gauge group.  

Here the electric charges of all particles are given by the formula
\begin{equation}
    Q = T_{3L} + T_{3R}+\frac{X}{2}
\end{equation}
where $T_{3L}$ and $T_{3R}$ are the third components of $SU(2)_L$ and $SU(2)_R$ respectively. Thus $(P,\,N,\,E)$ have $Q=(2/3,\,-1/3,\,-1)$, which enables mixing of these vector-like fermions with the up-type quarks, down-type quarks and charged leptons, respectively. Such mixings are responsible for the generation of ordinary quark and lepton masses in this scenario. Note the absence of an electrically neutral vector-like lepton, which is crucial for realizing Dirac neutrinos in the framework.  Such vector-like leptons with $Q=0$ are not part of the simplest anomaly-free representations of $SU(5)_L \times SU(5)_R$.

We have denoted the $U(1)$ factor of the gauge symmetry as $U(1)_X$.  This $U(1)$ may be identified as $(B-L)$, as can be seen from the charge assignment of Eq. (\ref{eq:fermion}) of the standard fermions. However, with this identification, $(B-L)$ would appear to be broken once the up-type quarks mix with the $P$ quark carrying a different $(B-L)$ charge (and similarly in the down-quark and charged lepton sectors).  There is an unbroken $(B-L)$ symmetry in the theory even after spontaneous gauge symmetry breaking, under which the $(P,\,N)$  quarks have charge $1/3$ and $E$ has charge $-1$, and the Higgs scalars have charge zero. We prefer to call this unbroken $U(1)$ as $(B-L)$, which may even be potentially  gauged, owing to its anomaly-free nature. In fact, we propose to gauge this $(B-L)$ in order to control Planck-induced Majorana masses of $\nu_L$ and $\nu_R$, as discussed further in Sec. \ref{sec:sec7}.

The Higgs sector of the model is quite simple, consisting of an $SU(2)_L$ doublet and an $SU(2)_R$ doublet: 
\begin{align}
\chi_L\ (1,2,1,1)& = \begin{pmatrix}
\chi_L^+ \\
 \chi_L^0 \\
\end{pmatrix}, \hspace{5mm}
\chi_R\ (1,1,2,1) = \begin{pmatrix}
 \chi_R^+ \\
 \chi_R^0 \\
\end{pmatrix} , 
\label{eq:Hspec}
\end{align}
Under parity, $\chi_L \leftrightarrow \chi_R$. Once the neutral components of these fields acquire vacuum expectation values (VEV) denoted as
\begin{align}
\langle \chi_L \rangle = 
\begin{pmatrix}
 0 \\
  \kappa_L \\
\end{pmatrix} \, , \hspace{7mm} 
\langle \chi_R \rangle = 
 \begin{pmatrix}
 0 \\
 \kappa_R \\
\end{pmatrix} \, 
\label{eq:vev}
\end{align}
the gauge symmetry breaks down to $SU(3)_c \times U(1)_{\rm em}$.  A hierarchy of VEVs is necessary with $\kappa_L \ll \kappa_R$, so that in the first stage of symmetry breaking the left-right symmetric gauge group breaks down to the standard model symmetry. This hierarchy can be realized with a Higgs potential that breaks parity symmetry softly via dimension-two mass terms:
\begin{eqnarray}
    V = -(\mu_L^2 \chi_L^\dagger \chi_L +\mu_R^2 \chi_R^\dagger \chi_R) + \frac{\lambda_1}{2}\left\{ (\chi_L^\dagger \chi_L)^2 + (\chi_R^\dagger \chi_R)^2 \right\}+ \lambda_2 (\chi_L^\dagger \chi_L)(\chi_R^\dagger \chi_R)~.
    \label{VEV}
\end{eqnarray}
Soft breaking of parity symmetry occurs when $\mu_L^2 \neq \mu_R^2$. We shall assume that this is the only source of explicit $P$-violation. $\mu_L^2 \neq \mu_R^2$ is necessary for phenomenology when the $SU(2)_R$ breaking scale is in the multi-TeV range, otherwise minimization of the tree-level Higgs potential  would lead to $\kappa_L = \kappa_R$, or $\kappa_L \kappa_R = 0$, neither of which is acceptable phenomenologically.  When the $SU(2)_R$ breaking scale is of order $10^{12}$ GeV, it has been shown that such soft parity breaking is not required for consistent phenomenology \cite{Hall:2018let}. 

Note that by separate $SU(2)_L$ and $SU(2)_R$ gauge rotations the VEVs $\kappa_L$ and $\kappa_R$ of Eq. (\ref{VEV}) can be made real, which plays in important role in solving the strong CP problem with parity symmetry.  This is in contrast to the conventional left-right symmetric theories where the VEVs of the bidoublet scalar used cannot be made real.

The masses of the  charged $W_{L,R}^\pm$ gauge bosons are given by
\begin{equation}
    M_{W_{L,R}^\pm}^2 = \frac{1}{2} g^2 \kappa_{L,R}^2 \, 
    \label{eq:gaugemass}
\end{equation}
with no tree-level mixing among the two. Here $g_L = g_R \equiv g$ has been assumed for the two $SU(2)$ gauge couplings, owing to parity symmetry. The neutral gauge boson mixing matrix can be found in Ref. \cite{Babu:2018vrl}, and will not be needed for our analysis. 

The most general Yukawa interactions of quark and leptons with the Higgs fields of the model are given by
\begin{align}
    \mathcal{L} &= y_u\ (\bar{Q}_L \tilde{\chi}_L + \bar{Q}_R \tilde{\chi}_R) P + y_d\ (\bar{Q}_L \chi_L + \bar{Q}_R \chi_R) N \nonumber\\
    &+ y_\ell\ (\bar{\Psi}_L \chi_L + \bar{\Psi}_R \chi_R) E + h.c. 
\end{align}
where $\tilde{\chi}_{L,R} = i \tau_2 \chi^*_{L,R}$. Here we have imposed parity symmetry with $Q_L \leftrightarrow Q_R$, $\Psi_L \leftrightarrow \Psi_R$, $P_L \leftrightarrow P_R$, $N_L \leftrightarrow N_R$, $E_L \leftrightarrow E_R$ and $\chi_L \leftrightarrow \chi_R$,  which relates various Yukawa coupling matrices.  
In addition, bare mass terms for the vector-like  fermions are allowed:
\begin{equation}
    \mathcal{L}_{\rm mass} = M_{p^0}\ \bar{P}P + M_{N^0}\ \bar{N} N + M_{E^0}\ \bar{E} E \, .
\end{equation}
Parity symmetry implies that $M_{P^0}$, $M_{N^0}$, and $M_{E^0}$ are all hermitian matrices.  
Thus, the $ 6 \times 6$ fermion mass matrices for the up- and down-type quarks ($M_u$ and $M_d$), and charged leptons ($M_\ell$) are given by (in a notation where $(\overline{u}_L,~\overline{P}_L$) multiplies the mass matrix $M_u$ from the left and $(u_R,~P_R)^T$ multiplies it from the right, and so forth):
\begin{align}
    M_u = 
    \begin{pmatrix}
    0 & y_u \,\kappa_L \\
    y_u^\dagger\, \kappa_R & M_{p^0}
    \end{pmatrix} \, , \hspace{5mm}
    M_d = 
    \begin{pmatrix}
    0 & y_d \,\kappa_L \\
    y_d^\dagger\, \kappa_R & M_{N^0}
    \end{pmatrix} \, ,  \hspace{5mm}
       M_\ell = 
    \begin{pmatrix}
    0 & y_\ell \,\kappa_L \\
    y_\ell^\dagger \,\kappa_R &  M_{E^0}
    \end{pmatrix} \, .
    \label{eq:chargedM}
\end{align}

This form of the quark mass matrices has the virtue that parity symmetry alone can solve the strong CP problem, without the  need for an axion.  Note that $\theta_{QCD} = 0$ due to parity, and from Eq. (\ref{eq:chargedM}) it follows that Arg\{Det$(M_uM_d)$\} = 0, which implies $\overline{\theta} = 0$ at tree level.  It has been shown in Ref. \cite{Babu:1989rb} that the one-loop corrections to $\overline{\theta}$ also vanish within this model. A small nonzero $\overline{\theta}$ will be induced at the two-loop level, which turns out to be sufficiently small to be consistent with the experimental limits on the neutron electric dipole moment arising from the strong CP parameter for most of the parameter space of the model.  

The mass matrices of Eq. (\ref{eq:chargedM}) can be block-diagonalized in the approximation $M^0 \gg y \kappa_{R} \gg y \kappa_L$ to obtain the following light fermion mass matrices:
\begin{equation}
    m_{u,d} = (y_{u,d}\ M_{P^0,N^0}^{-1}\ y_{u,d}^\dagger)\, \kappa_L \kappa_R\, , \hspace{5mm}   m_\ell = (y_\ell M_{E^0}^{-1} y_\ell^\dagger) \, \kappa_L \kappa_R \, .
    \label{eq:fermionmass}
\end{equation}
This is the generalized seesaw mechanism for charged fermion masses in this framework.  If one ignores family mixing, these mass formulas become
\begin{equation}
    m_{u_i} \approx \frac{ y_{u_i}^2 \kappa_L \kappa_R}{M_{P_i^0}},~~~~
        m_{d_i} \approx \frac{ y_{d_i}^2 \kappa_L \kappa_R}{M_{N_i^0}},~~~~
            m_{\ell_i} \approx \frac{ y_{\ell_i}^2 \kappa_L \kappa_R}{M_{E_i^0}},
            \label{approx}
\end{equation}
which shows the light fermion masses depending quadratically on the Yukawa couplings.  As a result, a milder hierarchy in the Yukawa couplings ($y_i = 10^{-3} - 1)$ is sufficient to explain the observed mass hierarchies as compared to the range $(10^{-6} - 1)$  needed in the standard model. We shall use the expressions of Eq. (\ref{eq:fermionmass}), rather than the crude approximations of Eq. (\ref{approx}) in our numerical study.

The approximation $y_i \kappa_R \ll M^0$ is very good for all the charged fermion masses, except for the top quark.  Ignoring the mixings of the top quark with other families, one can write more exact expressions for the physical top-partner mass (which we shall denote simply as $M_P$) and the top quark mass:
\begin{equation}
    m_t \, M_P = y_t^2 \,\kappa_L \kappa_R,~~~M_P = \sqrt{M_{P^0}^2+ |y_t \kappa_R|^2}~.
\end{equation}
This suggests the definition of a parameter $r$ with its range $0 \leq r \leq1$ as follows:
\begin{equation}
    M_{P^0} = r\ M_P, \hspace{10mm} r= \sqrt{1- \left|\frac{m_t}{M_p}\right| \frac{M_{W_R}}{M_{W_L}}} \, .
    \label{eq:baremp}
\end{equation}
We shall make use of Eq. (\ref{eq:baremp}) in our numerical study of the Dirac neutrino mass within these models. In the bottom quark sector we shall denote the bottom-partner mass to be simply $M_N$. (The other vector-like  quark masses do not enter our discussions of neutrino masses.)

Note that the neutrino masses remain zero at tree level in these models.  There is no vector-like neutral lepton that could have induced tree-level masses for the neutrinos. Neutrino masses do not remain zero at the quantum level, small Dirac
masses are generated via two-loop diagrams within this framework, which we discuss in the next section.


\section{Dirac Neutrino Masses as Two-loop Radiative Corrections }\label{sec:sec3}
\begin{figure}
\centering
    \includegraphics[scale=0.08]{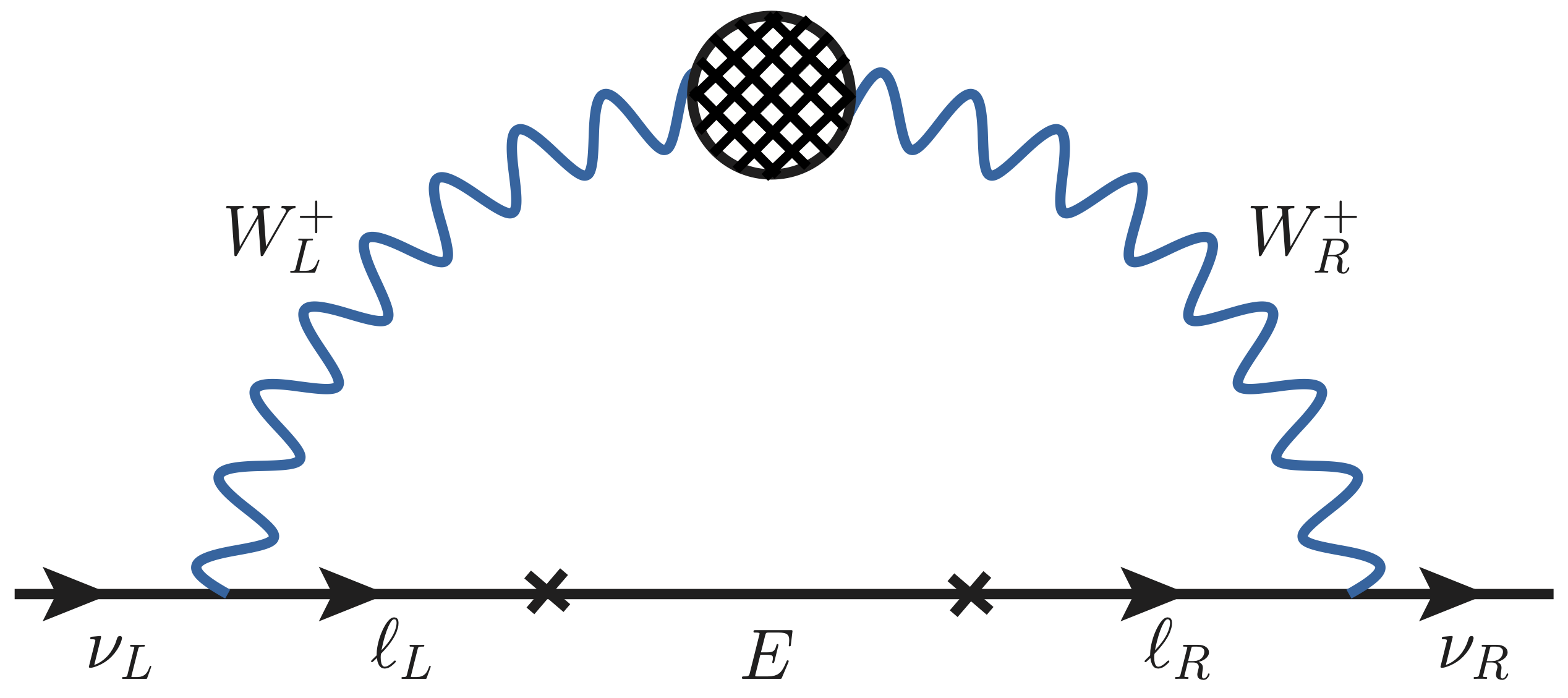}\hspace{10mm} \includegraphics[scale=0.5]{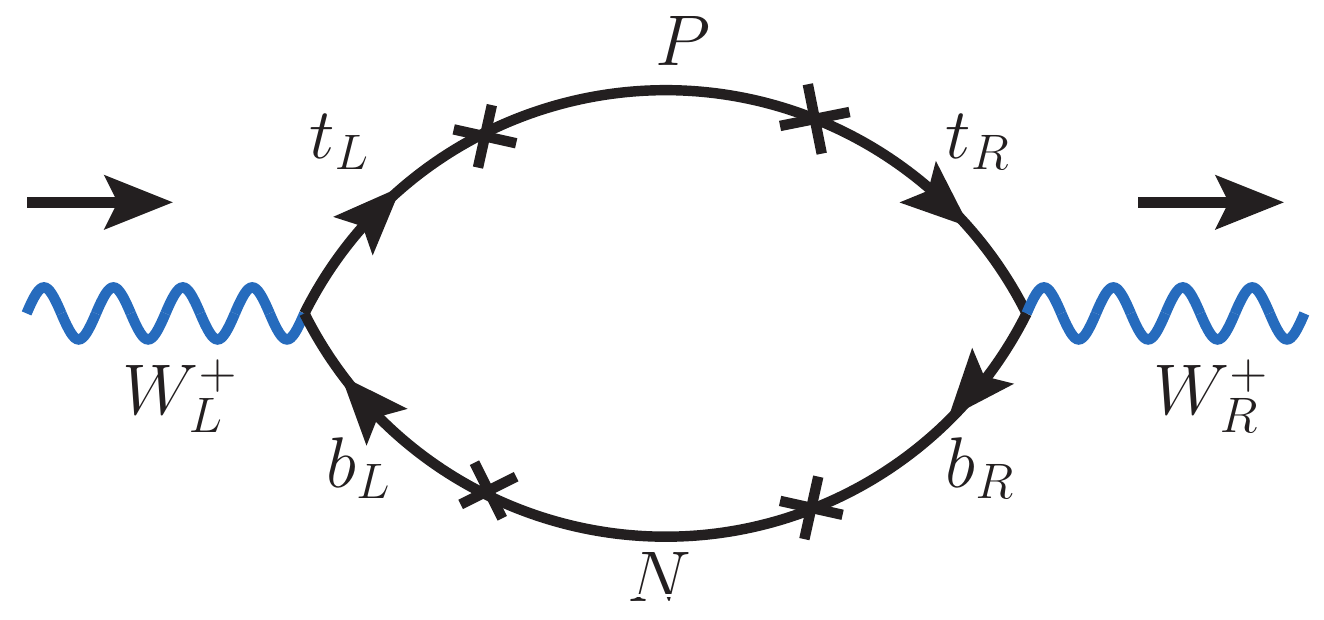}
    (a)~~~~~~~~~~~~~~~~~~~~~~~~~~~~~~~~~~~~~~~~~~~~~~~~~~~~~~~~~(b)
    \caption{Left: One-loop finite Dirac neutrino mass generated via the mixing shown in Fig.~(b). Right: One-loop diagram that induces mixing between $W_L^+ - W_R^+$.    }
    \label{fig:loop}
\end{figure}
The gauge bosons $W_L^\pm$ and $W_R^\pm$ do not mix at the tree level in this class of models which employs Higgs doublet for symmetry breaking. However, such a mixing, parameterized by an angle $\xi$, is induced at one-loop level as shown in the Fig.~\ref{fig:loop} (b) with internal top and bottom quarks. This mixing angle, which is finite, is approximately given by
\begin{equation}
    \xi \approx \frac{\alpha}{4 \pi \sin^2 \theta_W} \frac{m_b m_t}{M_{W_R}^2 } \, .
    \label{eq:mixing}
\end{equation}
Note that $\xi$ is sensitive to only the top and bottom quark masses, and not to the lighter quark masses.  
Once a nonzero $\xi$ is induced, the two-loop $W_L^\pm-W_R^\pm$ diagram of  Fig.~\ref{fig:loop} (a) would induce  small and finite Dirac masses for the neutrinos. The momentum of the $W_L^\pm$ and $W_R^\pm$ need to be kept nonzero -- unlike in the estimation of the $W_L^\pm-W_R^\pm$ mixing angle $\xi$. Keeping this in mind, the structure of the induced Dirac neutrino mass matrix is given by \cite{Babu:1988yq} 
\begin{equation}
    M_{\nu^D} = \frac{-g^4}{2} y_t^2 y_b^2 y_\ell^2 \kappa_L^3 \kappa_R^3 \frac{r\ M_{P} M_N M_{E_\ell}}{M_{W_L}^2 M_{W_R}^2}\ I_{E_\ell} \, ,
    \label{eq:Diracnu}
\end{equation}
where the parameter $r$ is defined in Eq.~\eqref{eq:baremp}.  Here we have worked in a basis where the vector-like lepton mass matrix is diagonal, but we allow for the light charged fermions to have a general mass matrix.  This is a consistent approach since we treat the light charged fermion masses perturbatively.  The flavor-dependent integrals $I_{E_\ell}$ of Eq. (\ref{eq:Diracnu}) (where $\ell = 1,2,3$ denote the heavy vector-like lepton flavor) are given by \cite{Babu:1988yq}
\begin{align}
    I_{E_\ell} &= \int \int \frac{d^4 k d^4 p}{(2\pi)^8} \frac{3 M_{W_L}^2 M_{W_R}^2 +  (p^2-M_{W_L}^2) (p^2-M_{W_R}^2)  }{k^2 (p+k)^2 (k^2-M_N^2) ((p+k)^2-M_p^2) p^2 (p^2-M_{E_\ell}^2) (p^2-M_{W_L}^2) (p^2-M_{W_R}^2)} \nonumber\\
    &= I_{1\ell} + I_{2\ell}
\end{align}
with
\begin{align}
    I_{1\ell} &= \int \int \frac{d^4 k d^4 p}{(2\pi)^8} \frac{3 M_{W_L}^2 M_{W_R}^2 }{k^2 (p+k)^2 (k^2-M_N^2) ((p+k)^2-M_p^2) p^2 (p^2-M_{E_\ell}^2) (p^2-M_{W_L}^2) (p^2-M_{W_R}^2)}~, \label{eq:I1} \\[4pt]
    I_{2\ell} &= \int \int \frac{d^4 k d^4 p}{(2\pi)^8} \frac{1  }{k^2 (p+k)^2 (k^2-M_N^2) ((p+k)^2-M_p^2) p^2 (p^2-M_{E_\ell}^2) }~. \label{eq:I2}
\end{align}
Here $M_P$ and $M_N$ are the physical masses of the top-quark partner and the bottom-quark partner respectively.

Now we turn to evaluating the loop integral $I_{1_\ell}$ and $I_{2_\ell}$ of Eq.~\eqref{eq:I1} and Eq.~\eqref{eq:I2}. 
For this purpose we define the following parameters which would enable us to express the integrals in terms of mass ratios:
\begin{equation}
     r_{1_\ell} = \frac{M_N^2}{M_{E_\ell}^2} \, , \hspace{10mm} r_{2_\ell} = \frac{M_P^2}{M_{E_\ell}^2} \, , \hspace{10mm} r_{3_\ell} = \frac{M_{W_R}^2}{M_{E_\ell}^2} \, , \hspace{10mm} r_{4_\ell} = \frac{M_{W_L}^2}{M_{E_\ell}^2} \, .
    \label{eq:ratio}
\end{equation}
We shall drop the subscript $\ell$ in these $r_{i_\ell}$, for brevity, but it should be kept in mind that these $r_i$ factors have flavor dependence (although ratios of $r_i$ are flavor-independent).   We follow the procedure outlined in Ref. \cite{vanderBij:1983bw, Broadhurst:1987ei, Ghinculov:1994sd, Usyukina:1994eg, McDonald:2003zj, Babu:2020bgz} to evaluate these integrals analytically. We first simplify these integrals by partial fractions \cite{tHooft:1972tcz} and performs integration by parts to obtain the analytical results.  
The total integral $I_E$ then can be written down as
\begin{align}
    I_E  &= I_1 + I_2 \nonumber\\
    &= \frac{1}{(16\pi^2)^2} \frac{1}{ M_N^2 M_P^2} \left[G_1(r_1,r_2,r_3,r_4) + G_2(r_1,r_2) \right] \, ,
    \label{eq:totI}
\end{align}
where the functions $G_1$ and $G_2$ correspond to integrals $I_1$ and $I_2$, respectively. The loop function $G_1(r_1, r_2, r_3, r_4)$ is found to be
\vspace{-4mm}
\begin{equation}
\begin{aligned}
    G_1( r_1,  & r_2, r_3, r_4) = \frac{3}{(r_3-1)(r_4-1)(r_4-r_3)}\Bigg[-\frac{\pi^2}{6} (r_1+r_2) (r_3-1) (r_3-r_4) (r_4-1)  \\
    &+ r_3 r_4(r_4-r_3)\left(r_1 F\left[\frac{1}{r_1}, \frac{r_2}{r_1}\right]+r_2 F\left[\frac{1}{r_2}, \frac{r_1}{r_2}\right]+ F\left[r_1, r_2\right]\right) \\
    &- (r_4-1) r_4\left(r_1 F\left[\frac{r_3}{r_1}, \frac{r_2}{r_1}\right]+r_2 F\left[\frac{r_3}{r_2}, \frac{r_1}{r_2}\right]+r_3 F\left[\frac{r_1}{r_3}, \frac{r_2}{r_3}\right]\right) \\
    &+ (r_3-1) r_3\left(r_1 F\left[\frac{r_4}{r_1}, \frac{r_2}{r_1}\right]+r_2 F\left[\frac{r_4}{r_2}, \frac{r_1}{r_2}\right]+r_4 F\left[\frac{r_1}{r_4}, \frac{r_2}{r_4}\right]\right) \\
    &+ (r_3 - r_4) (r_3-1) (r_4-1) \left( r_2 Li_2\left[1-\frac{r_1}{r_2}\right] + r_1 Li_2\left[1-\frac{r_2}{r_1}\right]\right) \\
    &+ r_3 r_4 ( r_3-r_4) \left( Li_2 [1-r_1]+ Li_2[1-r_2]  + r_1 Li_2\left[\frac{r_1-1}{r_1}\right] + r_2 Li_2 \left[\frac{r_2-1}{r_2}\right] \right) \\
    &+ r_4 (r_4-1)  \left(  r_3 Li_2\left[1-\frac{r_1}{r_3}\right] +  r_3 Li_2\left[ 1 - \frac{r_2}{r_3} \right]  + r_1 Li_2[1-\frac{r_3}{r_1}] + r_2 Li_2[1-\frac{r_3}{r_2}] \right) \\
     &- r_3 (r_3-1)  \left( r_4 Li_2\left[1-\frac{r_1}{r_4}\right] + r_4 Li_2\left[ 1 - \frac{r_2}{r_4} \right] + r_1 Li_2[1-\frac{r_4}{r_1}] + r_2 Li_2[1-\frac{r_4}{r_2}] \right) 
    \Bigg] \, .
\end{aligned}
\end{equation}
And the loop function $G_2(r_1, r_2)$ is evaluated to be
\begin{align}
    G_2(r_{1},r_2) &= -\frac{\pi^2}{6} (r_1 + r_2 + 1) -  F\left[r_1, r_2\right]+ Li_2\left[ 1-r_1\right] + Li_2\left[1-r_2\right]   \nonumber\\
    &~~~+ r_1 \left( -F\left[\frac{1}{r_1}, \frac{r_2}{r_1}\right]+ Li_2\left[ 1-\frac{1}{r_1}\right] + Li_2\left[1- \frac{r_2}{r_1}\right]  \right) \nonumber\\ 
    &~~~+ r_2 \left( -F\left[\frac{1}{r_2}, \frac{r_1}{r_2}\right]+ Li_2\left[1- \frac{1}{r_2}\right] + Li_2\left[ 1-\frac{r_1}{r_2}\right]   \right) \, .
\end{align}
The dilogarithm function $\mathrm{Li}_2(x)$ is defined as
\begin{equation}
\mathrm{Li}_{2}(x)=-\int_{0}^{x} \frac{\log (1-y)}{y} d y \, ,
\end{equation}
and the function $F(a,b)$ is defined as
\begin{align} F(a,b)=&-\frac{1}{2} \log a \log b-\frac{1}{2}\left(\frac{a+b-1}{\sqrt{\Delta}}\right)\Bigg\{ \operatorname{Li}_{2}\left(\frac{-x_{2}}{y_{1}}\right)+\operatorname{Li}_{2}\left(\frac{-y_{2}}{x_{1}}\right)-\operatorname{Li}_{2}\left(\frac{-x_{1}}{y_{2}}\right) \nonumber\\
&-\operatorname{Li}_{2}\left(\frac{-y_{1}}{x_{2}}\right) +\operatorname{Li}_{2}\left(\frac{b-a}{x_{2}}\right)+\operatorname{Li}_{2}\left(\frac{a-b}{y_{2}}\right)-\operatorname{Li}_{2}\left(\frac{b-a}{x_{1}}\right)-\operatorname{Li}_{2}\left(\frac{a-b}{y_{1}}\right)\Bigg\} \, ,
\label{eq:D18}
\end{align} 
where $\Delta=\left(1-2(a+b)+(a-b)^{2}\right)$ and 
\begin{align}
x_{1}=\frac{1}{2}(1+b-a+\sqrt{\Delta }), &
\hspace{5mm} 
x_{2}=\frac{1}{2}(1+b-a-\sqrt{\Delta}) \, , \nonumber\\ 
y_{1}=\frac{1}{2}(1+a-b+\sqrt{\Delta}), & 
\hspace{5mm} {y_{2}=\frac{1}{2}(1+a-b-\sqrt{\Delta})} \, .
\end{align}
By explicit symmetrization under  $a \leftrightarrow b$ and using the relations
\begin{align}
{\rm Li}_{2}(1-z)&=-\mathrm{Li}_{2}(z)-\log z \log (1-z)+\frac{1}{6} \pi^{2} \, , \nonumber \\ 
{\rm Li}_{2}\left(\frac{1}{z}\right) &=-\mathrm{Li}_{2}(z)-\frac{1}{2} \log ^{2}(-z)-\frac{1}{6} \pi^{2} \, ,
\end{align}
the function $F(a,b)$ can be simplified to:
\begin{eqnarray}
    F(a,b)=-\frac{1}{2} \log a \log b &- & \left(\frac{a+b-1}{\sqrt{\Delta }}\right) \bigg\{ \mathrm{Li}_{2}\left(\frac{-x_2}{y_{1}}\right)+\mathrm{Li}_{2}\left(\frac{-y_{2}}{x_{1}}\right)+\frac{1}{4} \log ^{2} \frac{x_{2}}{y_{1}} \nonumber  \\
 &&+\frac{1}{4} \log ^{2} \frac{y_{2}}{x_{1}} +\frac{1}{4} \log ^{2} \frac{x_{1}}{y_{1}}-\frac{1}{4} \log ^{2} \frac{x_{2}}{y_{2}}+ \frac{\pi^2}{6}\bigg\} \, .
 \label{eq:D16}
\end{eqnarray}
It is to be noted that function $F(a,b)$ in Eq.~(\ref{eq:D16}) can have a non-zero imaginary part \cite{Coleman:1965xm, McDonald:2003zj}. However, these  imaginary components cancel out with judicious logarithmic branch cut choice. The real part of Eq.~(\ref{eq:D16}) is in full agreement with Eq.~\eqref{eq:D18}.

\subsection{Evaluation of the loop integrals}
\begin{figure}[!t]
    \includegraphics[height=6.3cm, width=7.7 cm]{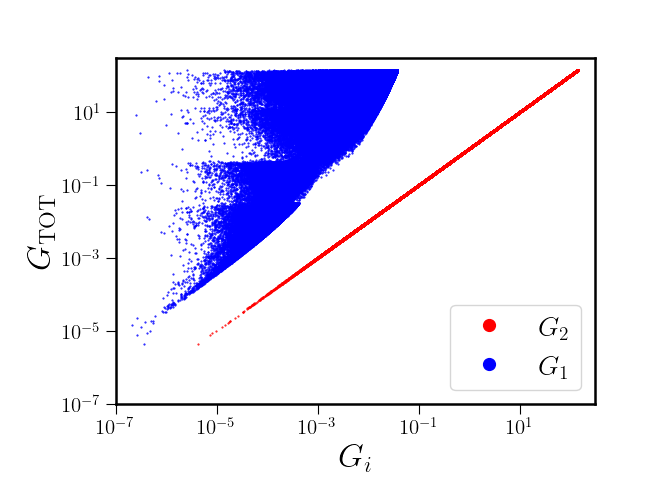}
    \includegraphics[height=5.5cm, width=7 cm]{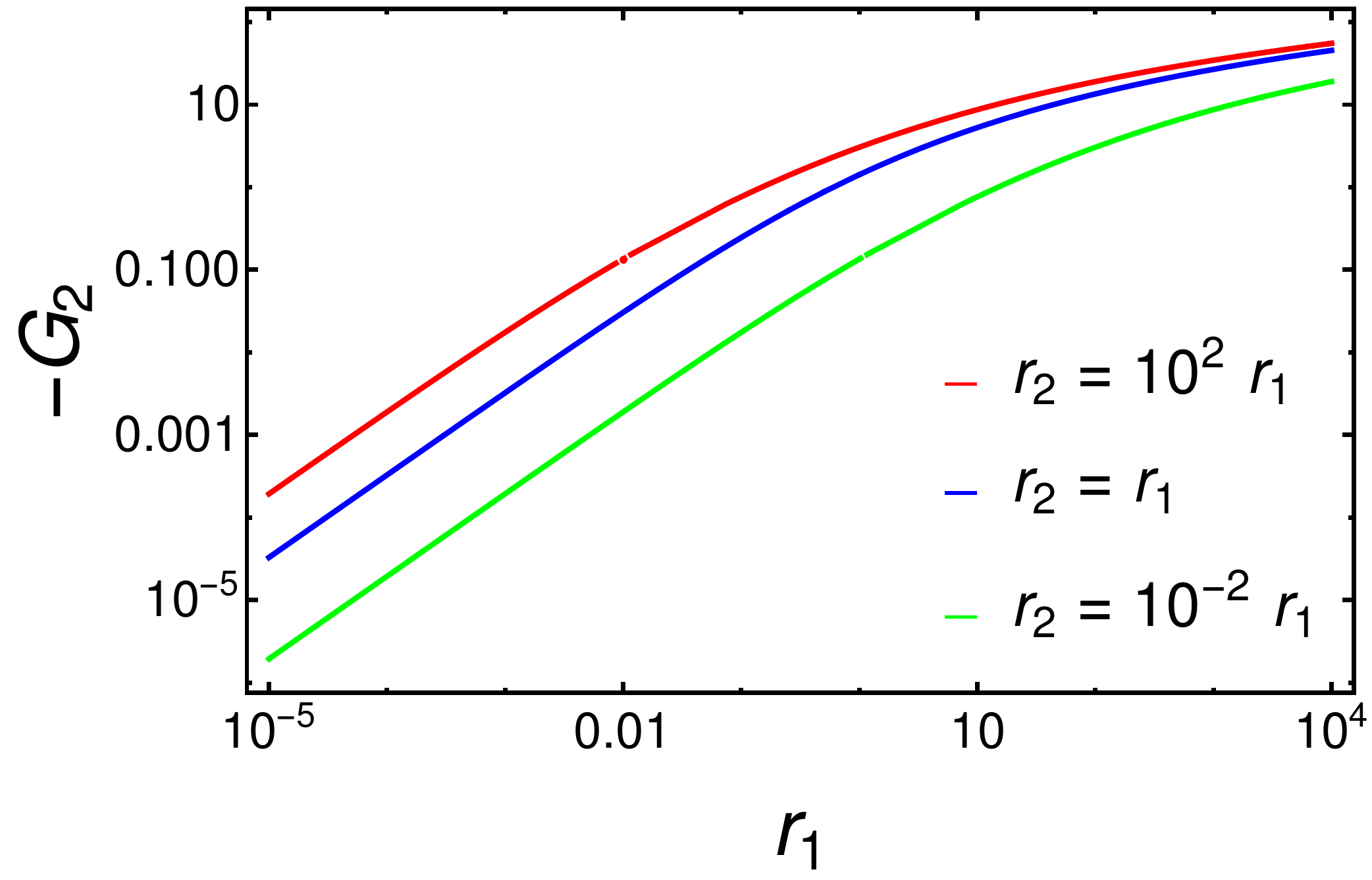}
    \caption{Left: Integral function $G_{\rm TOT} = G_1 + G_{2}$ vs $G_i\ (i=1,2)$. Right: contribution from $G_2(r_1, r_2)$ as a function of mass ratio $r_1$ for three different choices $r_2=100\ r_1$ (red), $r_2= r_1$ (blue), and $r_2=0.01\ r_1$ (green). }
    \label{fig:integral}
\end{figure}
\begin{figure}
\centering
    \includegraphics[scale=0.07]{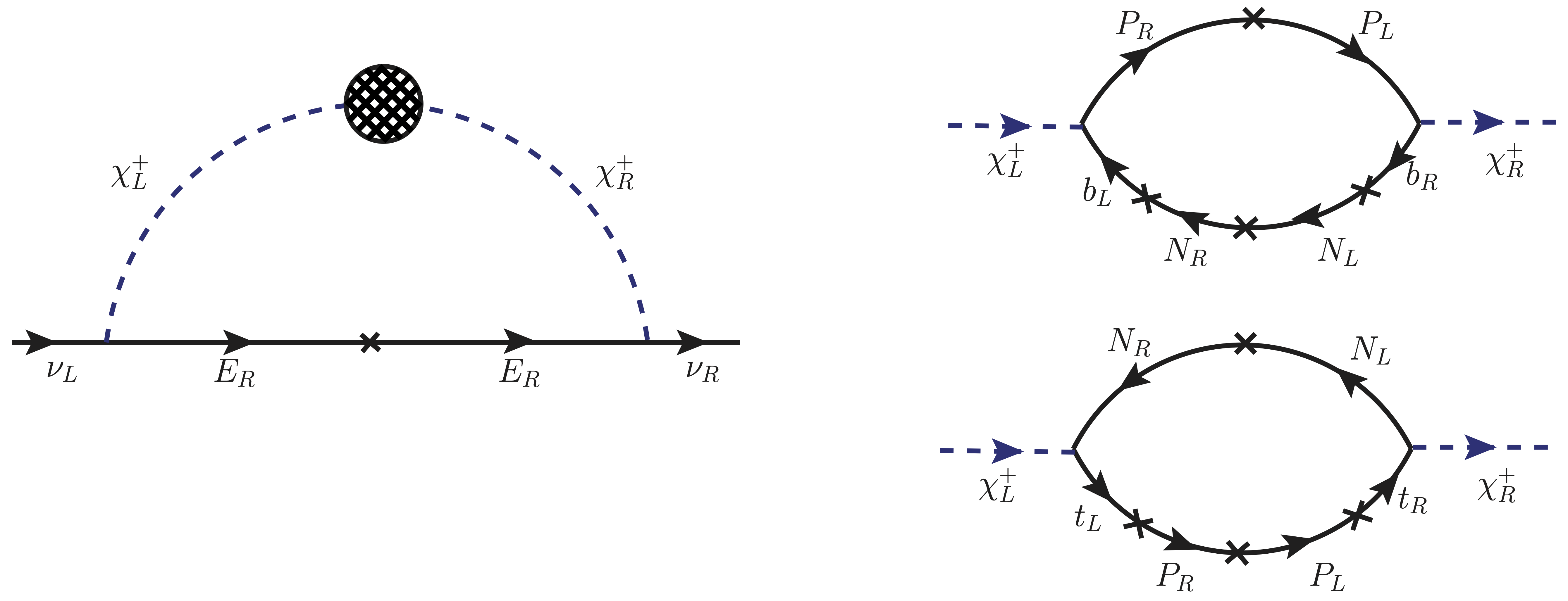}
    \caption{The longitudinal $W_{L,R}^\pm$ gauge boson contributions to the neutrino mass, expressed in terms of the Goldstone boson contributions. }
    \label{fig:loop2}
\end{figure}

After scanning the parameters of the model numerically imposing the phenomenolgical constraints that $M_{W_R^\pm} \geq 4$ TeV, and $0\leq r \leq 1$ in Eq. (\ref{eq:baremp}), we find that the function $G_2$ dominates $G_1$ over the entire the parameter space. This is  shown in Fig.~\ref{fig:integral} (left panel), where we have plotted $G_{\rm TOT} = G_1+G_2$ as a function of  $G_1$ and $G_2$. The blue and the red regions in Fig.~\ref{fig:integral} respectively represents $G_1$ and $G_2$. For any choice of parameters the total contribution $G_{\rm TOT}$ is approximately equal to $G_2$. This feature is not surprising, as $G_2$  arises from the exchange of longitudinal $W_{L,R}^\pm$, while $G_1$ corresponds to contribution from the transverse modes.  The Goldstone boson contributions are explicitly shown in Fig. \ref{fig:loop2}. These contributions to the neutrino mass do not decouple when the $W_R$ mass is taken to infinity, as can be confirmed by the power counting of Fig. \ref{fig:loop2}.  This limit should be accompanied by taking the vector-like fermion masses to infinity as well, so that the light fermion masses remain finite.  The transverse $W_{L,R}$ contributions to the neutrino masses, on the other hand, would decouple as $W_R$ mass is taken to infinity, thus leading to the result $G_1 \ll G_2$. We therefore safely ignore the contributions from $G_1$ in our numerical study. Fig.~\ref{fig:integral} (right panel) shows the contribution from $G_2(r_1, r_2)$ as a function of mass ratio $r_1$ for three different choices of $r_2$: $r_2=100\ r_1$ (red), $r_2= r_1$ (blue), and $r_2=0.01\ r_1$ (green). We also note that our results agree with a Feynman parametric integral representations given in  Ref.~\cite{Babu:1988yq}, upon numerically integrating those functions.

It is useful to find the asymptotic behavior of the function $G_2(r_1, r_2)$. Following the approximations developed in  Ref.~\cite{vanderBij:1983bw}, we find the leading asymptotic behavior of $G_2$ in various limits to be:
\begin{equation}
G_2(r_1, r_2)\ \rightarrow\  
\left\{\begin{array}{ll}
\frac{-\pi^2}{6} (r_1 + r_2 - 2 r_1 r_2) - r_1 r_2(1 - \log r_1 \log r_2)   \\
~~~~~~~~~~~~~~~~~+ r_2 Li_2\left[ 1-\frac{r_1}{r_2}\right] + r_1 Li_2\left[ 1-\frac{r_2}{r_1}\right]  &\hspace{8mm} \text { for } \ r_{1}, r_2 \ll 1 \, ,\\[10pt] 
\frac{-\pi^2}{6} (1 + r_2) +  Li_2\left[1-r_2\right]  +  r_2\ Li_2\left[1- \frac{1}{r_2}\right]  &\hspace{8mm} \text { for } \  r_{2} \ll  r_1 \gg 1 \, ,\\[10pt]
 -(2 + \frac{1}{4}  \log r + \frac{1}{2} \log^2 r) &\hspace{8mm} \text { for } \  r_{1} \simeq r_2 = r \gg  1  .
\end{array}\right.
\label{eq:asym2}
\end{equation}
Furthermore, in the limit of $r_1, r_2 \ll 1$ and $r_2 \ll r_1$, the first expression given in Eq.~\eqref{eq:asym2} further simplifies to
\begin{equation}
    G_2 (r_1, r_2) \simeq r_2 \left( \frac{\pi^2}{3}(r_1-1) - (r_1+1) + r_1 \log r_1 \log r_2 + \log\frac{r_2}{r_1} -\frac{1}{2} \log^2 r_2)\right).
\end{equation}
For $r_1 \ll r_2$, the approximation for the  function can be obtained by simply replacing $r_1 \leftrightarrow r_2$. Similarly, in the limit $r_2 \ll r_1 \gg 1$ and $r_2 \ll 1$, the second expression of Eq.~\eqref{eq:asym2} reduces to 
\begin{equation}
    G_2 (r_1, r_2) \simeq -r_2 \left( 1+ \frac{\pi^2}{3} + \frac{1}{2} \log r_2 (-1 + \log r_2)  \right) \, .
\end{equation}

\section{Fits to Neutrino Oscillation Data}
\label{sec:sec4}

\begin{table}[t!]
{\footnotesize
\centering
\begin{tabular}{|c|c|c|c|c|c|}
\hline \hline
 {\textbf{Oscillation}}   &   {\textbf{$3\sigma$ range}} &  \multicolumn{4}{c|}{\textbf{Model prediction}}\\
 \cline{3-6}
 {\bf parameters} & \bf{{\tt NuFit5.1}}~\cite{Esteban:2020cvm} & BP I (NH) &  BP II (NH) & BP III (IH) &  BP IV (IH) \\ \hline \hline
 $\Delta m_{21}^2 (10^{-5}~{\rm eV}^2$)  &   6.82 - 8.04   &  7.42 & 7.38 &  7.35 &  7.35  \\ \hline
 $\Delta m_{23}^2 (10^{-3}~{\rm eV}^2) $(IH) &   2.410 - 2.574  & - &  - & 2.48 & 2.52 \\ 
 $\Delta m_{31}^2 (10^{-3}~{\rm eV}^2) $(NH) &   2.43 - 2.593   & 2.49 & 2.51 & - & - \\ \hline
  $\sin^2{\theta_{12}}$   &   0.269 - 0.343 &   0.324 &  0.301 & 0.306 & 0.310 \\ \hline
 $\sin^2{\theta_{23}}$  (IH) &   0.410 - 0.613  &  -  &  - & 0.510 & 0.550 \\
 $\sin^2{\theta_{23}}$ (NH)  &   0.408 - 0.603  &  0.491  &  0.533 & - & - \\ \hline 
  $\sin^2{\theta_{13}} $  (IH) &   0.02055 - 0.02457  &  -  &  - & 0.0219 & 0.0213 \\
  $\sin^2{\theta_{13}}  $(NH)  &   0.02060 - 0.02435  &  0.0234  & 0.0213 & - & - \\ \hline
  $\delta_{\rm CP}$ (IH) & 192 - 361 & - & - & $236^\circ$ & $279^\circ$  \\
  $\delta_{\rm CP}$ (NH) & 105 - 405 & $199^\circ$ & $280^\circ$ & - & - \\ 
  \hline
 \multicolumn{2}{|c|}{$m_{\rm light}$ $(10^{-3})$ eV}  & 0.66 & 2.04 & 14.1 & 8.50 \\  \hline  \hline
\multicolumn{2}{|c|}{$M_{E_1}/M_{W_R}$}  & 917 & 45.5 & 1936 & 1990 \\  \hline 
\multicolumn{2}{|c|}{$M_{E_2}/M_{W_R}$} & 0.650 & 0.43 & 0.12 & 0.11 \\  \hline 
\multicolumn{2}{|c|}{$M_{E_3}/M_{W_R}$}   & 0.019 & 0.029  & 0.015  & 0.012 \\  \hline \hline
\end{tabular}
\caption{Fits to the neutrino oscillation parameters in the model with normal and inverted hierarchy. For comparison, the $3\sigma$ allowed range for the oscillation parameters are also given.}
\label{tab:fit}
}
\end{table}

\begin{table}[!t]
{\footnotesize
    \centering
    \begin{tabular}{|c|c|c|c|c|c|c|c|c|c|c|}
    \hline \hline
          & $M_N/M_{W_R}$ & $y_1$ & $y_2$ & $y_3$ & $\phi_{12}$ & $\phi_{13}$ & $\phi_{23}$ & $\varphi$ & r ($10^{-3}$ )  \\
         \hline
        {\tt BP I} (NH)    & 50.4 & 0.0051 & 0.012 & 0.076 & 23.8$^\circ$ & 31.5$^\circ$ & 12.4$^\circ$ & 80.2$^\circ$ & 0.174  \\
        \hline
        {\tt BP II} (NH)  & 59.9 & 0.02 & 0.005 & 0.01 & 67.2$^\circ$ & 49.0$^\circ$ & 20.3$^\circ$ & 87.1$^\circ$ & $0.264$ \\ 
        \hline
        {\tt BP III} (IH)  & 41.5 & 0.052 & 0.0053 & 0.0095 & 4.59$^\circ$ & 4.25$^\circ$ & 17.2$^\circ$ & 5.79$^\circ$ & 2.17 \\
        \hline
        {\tt BP IV} (IH)   & 41.4 & 0.053 & 0.0047  & 0.0091 & 5.2$^\circ$ & 4.8$^\circ$ & 21.1$^\circ$ & 6.28$^\circ$ & 2.14 \\
          \hline \hline
    \end{tabular}
    \caption{Model parameters that give neutrino oscillation fit of Table~\ref{tab:fit}. Here $M_P/M_{W_R} \simeq 2.0$. }
    \label{tab:input}
}
\end{table}

In this section, we show that the model can correctly reproduce the neutrino oscillation data by fitting the model parameters to the observables  ($\Delta m_{21}^2,\,\Delta m_{31}^2, \,\sin^2 \theta_{13},\, \sin^2 \theta_{23}, \,\sin^2 \theta_{12}$). We find good fits to the normal ordering of neutrino masses as well as inverted ordering. Furthermore, the model does not place any restriction of the CP violating phase in neutrino oscillation, as we find fits for any value of $\delta_{CP}$. 

The Dirac neutrino mass matrix arising from the two-loop diagrams  has a structure given in Eq.~\eqref{eq:Diracnu}. Upon evaluation of the loop integral $I_E$, this matrix can be written as 
\begin{equation}
   M_{\nu^D} =   y_\ell\ M_E I_E \ y_\ell^\dagger   \, ,
    \label{eq:nuM}
\end{equation}
where $I_E$ is a dimensionless function that contains all the relevant factors and the loop functions:
\begin{equation}
    I_E =   -\frac{g^2}{(16 \pi^2)^2} \frac{m_t m_b}{M_{W_L} M_{W_R}}\ r\ \left[G_1(r_1,r_2,r_3,r_4) + G_2(r_1,r_2) \right] \, ,
    \label{eq:IE}
\end{equation}
In obtaining the form of Eq. (\ref{eq:IE})
we have used  Eq.~\eqref{eq:gaugemass},  Eq.~\eqref{eq:fermionmass}, as well as Eq. (\ref{eq:baremp}).

The charged lepton mass matrix is given by
\begin{align}
     m_\ell &= (y_\ell\, M_{E^0}^{-1}\, y_\ell^\dagger) \,\kappa_L \kappa_R \,.
     \label{eq:chargL}
\end{align}
We wish to stay in a basis where the charged lepton mass matrix is diagonal.  For this purpose, we note that the heavy lepton mass matrix $M_{E^0}$ can be taken to be diagonal and real without loss of generality.  In the seesaw approximation, the matrix is very nearly equal to the physical vector-like lepton mass matrix $M_E = {\rm diag}(M_{E_1}, \, M_{E_2},\, M_{E_3})$.  We can also parametrize $y_\ell$ in the same basis as
\begin{equation}
    y_\ell = y_D \, V^\dagger
\end{equation}
where $V$ is some unitary matrix and $y_D = {\rm diag}(y_1, \,y_2,\, y_3)$.  The hermitian matrix $m_\ell$ of Eq. (\ref{eq:chargL}) can be diagonalized to obtain the physical charged lepton mass matrix $M_\ell$ as
\begin{equation}
    m_\ell = U M_\ell U^\dagger
\end{equation}
with $M_\ell = {\rm diag}(m_e,\,m_\mu,\, m_\tau)$.  

We can now invert Eq. (\ref{eq:chargL}) and solve for the matrix $M_E$ (which at the leading order is the same as $M_{E^0}$). Doing so we obtain
\begin{equation}
    M_E = \frac{2 M_{W_L}M_{W_R}}{g^2} \left( V y_D U M_\ell^{-1} U^\dagger y_D V^\dagger \right) \, .
    \label{eq:heavylep}
\end{equation}
For the matrix $M_E$ to be diagonal, the unitary matrix $V$ appearing in Eq. (\ref{eq:heavylep}) will be determined once the other entries of this equation are specified.  In addition to $y_D = {\rm diag}(y_1, \,y_2,\, y_3)$, we must specify the elements of the unitary matrix $U$.  If we write $U = P.\hat{U}.Q$, where $P$ and $Q$ are diagonal phase matrices, $P$ can be absorbed into $V$ and $Q$ disappears from Eq. (\ref{eq:heavylep}).  This suggests that $U$ in Eq. (\ref{eq:heavylep}) can be parametrized in terms of three mixing angles and one phase, very much like the standard form of the CKM matrix.  We denote the three angles of $U$ to be ($\phi_{12}$, $\phi_{13}$, $\phi_{23}$) and the single phase of $U$ to be $\varphi$ in the standard CKM-like parametrization of  $U$. 
Thus, in a basis where the charged lepton mass matrix is diagonal, the Dirac neutrino mass matrix of Eq. (\ref{eq:nuM}) takes the form
\begin{equation}
     M_{\nu^D} = -\frac{2 m_t m_b}{(16\pi^2)^2}\ r \left( U^\dagger y_D^2 U M_\ell^{-1} U^\dagger y_D V^\dagger G_2(r_{1_\ell},r_{2_\ell}) V y_D U \right) \,. 
     \label{eq:numass}
\end{equation}
The neutrino mass matrix given by Eq.~\eqref{eq:numass} is diagonalized by a unitary transformation 
\begin{equation}
    U^T_{\text{PMNS}} M_\nu U_{\text{PNMS}} \ = \ \widehat{M}_\nu \, ,
    \label{eq:nudiag}
\end{equation}
where $\widehat{M}_\nu$ is the diagonal mass matrix and $U_{\text{PMNS}}$ is the $3\times3$ PMNS lepton mixing matrix. 

To summarize, we work in a basis where the charged lepton mass matrix is diagonal. The two-loop neutrino mass diagram is evaluated in a basis where the vector-like lepton mass matrix is diagonal.  The unitary matrix $U$ of Eq. (\ref{eq:numass}) is a function of $(\phi_{12},\, \phi_{13},\, \phi_{23},\, \varphi)$.  For any given choice of $y_D = {\rm diag}(y_1,\,y_2,\,y_3)$, $U$ and $M_{W_R}$, the eigenvalues of $M_E$ are determined via Eq. (\ref{eq:heavylep}). The unitary matrix $V$ is also determined as the diagonalizing matrix for $M_E$, since we work in a basis where the vector-like lepton mass matrix is diagonal.  Once $M_E$ eigenvalues are known, and once $M_P$ and $M_N$ are specified, the loop function $G_2(r_1,r_2)$ is also fixed. Thus the neutrino mass matrix is determined via Eq. (\ref{eq:numass}). The variables of the neutrino fit are three Yukawa couplings $(y_1,\,y_2,\,y_3)$, three mixing angles ($\phi_{12},\,\phi_{13},\,\phi_{23})$ and a phase $\varphi$ that parametrize the unitary matrix $U$, the parameter $r$, and the masses $ M_P,\,M_N$. Note that $M_{W_R}$ is not counted as a free parameter in this set, since it is fixed in terms of the parameter $r$, see Eq. (\ref{eq:baremp}). 
This is a total of 10 parameters that we shall vary in our numerical fits. 

We then perform a numerical scan over nine of these ten input parameters, keeping $M_P/M_{W_R} \simeq 2.0$ fixed. The best fit values of these parameters that give consistent fits to neutrino oscillations are given in 
Table.~\ref{tab:input}  for four benchmark points.  We have found good fits for both normal (NH) and inverted (IH) hierarchy of neutrino masses with the constrained minimization where the observables are required to be within $3\sigma$ of their experimentally measured values. Fits to both the NH and the IH are given in Table~\ref{tab:fit}, corresponding to the input parameter given in Table~\ref{tab:input}. These fits are in excellent agreement with the observed experimental values. 

As an independent check of the procedure we adopted, we show that the benchmark point BPI (and similarly the other benchmark points) correctly reproduces the SM-charged lepton masses ($m_e, m_\mu, m_\tau$) and neutrino masses and mixings given in Table~\ref{tab:fit}. Here we fix $M_{W_R} = 30$ TeV, $M_N= 1.51 \times 10^3$ TeV, and $M_P = 59.7$ TeV. This leads to the following values for the integral $I_E$, mass matrix $M_E$, and Yukawa matrix $y_\ell$:
\begin{align}
&~~~~~~~~~I_E = -{\rm diag}\ (1.25 \times 10^{-16}, 7.36 \times 10^{-12}, 4.80 \times 10^{-11}) \, , \nonumber \\
&~~~~~~~~~M_E = {\rm diag}\ (2.75 \times 10^{4}, 19.5, 0.57)\ {\rm TeV } \, , \nonumber \\
  y_\ell &= 10^{-2} \begin{pmatrix}
      0.058 i &~~  -0.063 - 0.041\ i &~~ -0.084+0.497 i \\
    0.036 -0.162 \ i &~~ -0.37 + 1.10\ i &~~ -0.18 + 0.033\ i \\
     7.48   &~~  1.12 &~~ -0.77  \\
\end{pmatrix} \, .
\end{align}
From these values, and using  Eq.~\eqref{eq:chargedM} we obtain the correct charged lepton masses and $6\times 6$ unitary matrix that diagonalizes $M_\ell$ of Eq.~\eqref{eq:chargedM}.  Furthermore, the Dirac neutrino mass matrix of Eq.~\eqref{eq:nuM} for this choice of input parameters is given by
\begin{equation}
 M_{\nu^D} =  10^{-1}\ \begin{pmatrix}
      0.0077 &~~  -0.0024 + 0.0098\ i &~~ -0.0084 - 0.0156\ i \\
     -0.0024 - 0.0098\ i &~~ 0.194 &~~ -0.055 + 0.173\ i \\
     -0.0084 + 0.0156\ i  &~~ -0.055 - 0.173\ i &~~ 0.39\\
\end{pmatrix} \,   {\rm eV}.
\label{eq:DiracM}
\end{equation}
The $3\times 3$ sub-block matrix, $U_\ell$, of the $6\times 6$ unitary matrix that diagonalizes $M_\ell$, together with the unitary matrix that diagonalizes $M_{\nu^D}$ of Eq.~\eqref{eq:DiracM} gives the desired PMNS matrix, $U_{\rm PMNS} = U_\ell^* U_\nu$, from which one obtains the mixing angles identical to those  given in Table \ref{tab:fit} corresponding to BPI. 

We conclude that the model provides excellent fits to neutrino oscillation data.  Both normal ordering and inverted ordering of neutrino masses  are permitted within the model. The CP violating phase can take a large range of values, as shown in Table \ref{tab:fit}. We have verified that $\delta_{\rm CP}$ in all four quadrants are admitted within this framework, both for NH and IH.

\section{Origin from \texorpdfstring{\boldmath{$SU(5)_L \times SU(5)_R$}}{SU5}}

\label{sec:sec5}

As noted earlier, the fermion spectrum of the model has a natural embedding in $SU(5)_L \times SU(5)_R$ unification \cite{Davidson:1987mh,Lee:2016wiy}.  Under this symmetry, all left-handed fermions of the SM fit into a ${\bf 10} + {\bf \overline{5}}$ of $SU(5)_L$ and all right-handed fermions of the SM fit into ${\bf 10} + {\bf \overline{5}}$ of $SU(5)_R$. The vector-like quarks and leptons fill these multiplets, with no additional degrees of freedom allowed.  Specifically, if we denote the vector-like fermions as $(U,\,D,\,E)$, their grouping is given by
\begin{equation}
   F_{L,R} = \left[\begin{array}{c}
D_{1}^{c} \\
D_{2}^{c} \\
D_{3}^{c} \\
e \\
-\nu
\end{array}\right]_{L,R} \, \hspace{10mm}
T_{L,R} =\frac{1}{\sqrt{2}}\left[\begin{array}{ccccc}
0 & U_{3}^{c} & -U_{2}^{c} & u_{1} & d_{1} \\
-U_{3}^{c} & 0 & U_{1}^{c} & u_{2} & d_{2} \\
U_{2}^{c} & -U_{1}^{c} & 0 & u_{3} & d_{3} \\
-u_{1} & -u_{2} & -u_{3} & 0 & E^{c} \\
-d_{1} & -d_{2} & -d_{3} & -E^{c} & 0
\end{array}\right]_{L,R} \, .
\end{equation}
Here the group transformation properties are
$F_L(\overline{5},1)$, $F_R(1,\overline{5})$, 
$T_L(10,1)$ and $T_R(1,10)$. Parity symmetry can be imposed on these multiplets under which $F_L \leftrightarrow F_R$ and $T_L \leftrightarrow T_R$. Note that there is no vector-like neutral lepton, which could have mixed with the neutrino.  This embedding shows the naturalness of Dirac neutrinos in the left-right symmetric models that we have developed here.

Although we shall not build full models based on $SU(5)_L \times SU(5)_R$ unification here, it is worth noting that the left-right symmetric models we have studied here arise as a natural intermediate symmetry in these unified theories.  The symmetry breaking chain can proceed as follows:
\begin{equation}
\begin{gathered}
S U(5)_{L} \times S U(5)_{R} \\
\downarrow \Lambda \\
S U(3)_{L} \times S U(2)_{L} \times U(1)_{L} \times S U(3)_{R} \times S U(2)_{R} \times U(1)_{R} \\
\downarrow \Lambda_{L R} \\
S U(3)_{c} \times S U(2)_{L} \times S U(2)_{R} \times U(1)_{L+R} \\
\downarrow \kappa_{R} \\
S U(3)_{c} \times S U(2)_{L} \times U(1)_{Y} \\
\downarrow \kappa_{L} \\
S U(3)_{c} \times U(1)_{\rm em}
\end{gathered}
\end{equation}
The gauge symmetry below the scale $\Lambda_{LR}$ can be identified as the left-right symmetry that we have studied.  We also note that the $U(1)$ symmetry realized at this scale is $U(1)_{L+R}$, which we have denoted as $U(1)_X$.  In principle, without enlarging the fermion spectrum, one could also gauge the orthogonal $U(1)$ in the left-right symmetric models, which would commute with the rest of the gauge symmetry.

\section{Cosmological Tests of the Model in \texorpdfstring{\boldmath{$N_{\rm eff}$}}{Neff}}
\label{sec:sec6}


\begin{figure}[!t]
\centering
    \includegraphics[width=0.85\textwidth]{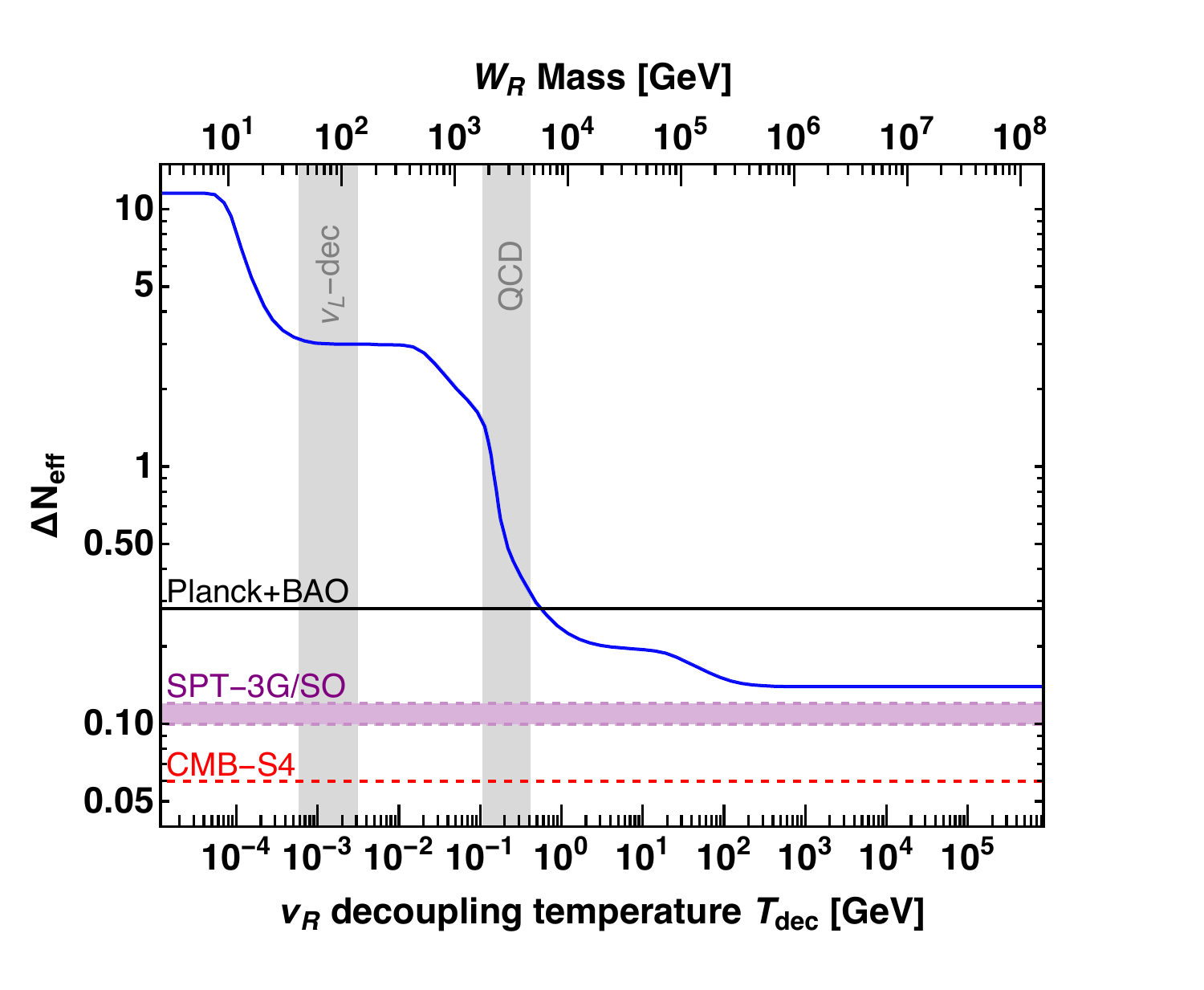}
    \caption{Contributions of three right-handed neutrinos to $\Delta N_{\rm eff}$ as a function of decoupling temperature $T_{\rm dec}$ (lower horizontal scale) as well as the $W_R$ mass (upper scale). The solid horizontal lines represents the current $2\sigma$ limit from Planck+ BAO \cite{Planck:2018vyg},  The future sensitivity reach of SPT-3G \cite{SPT-3G:2014dbx}, SO \cite{SimonsObservatory:2019qwx} and CMB-S4 \cite{Lee:2013bxd,Abazajian:2019eic} are indicated.  }
    \label{fig:CMB}
\end{figure}

If neutrinoless double beta decay is observed experimentally, it would establish lepton number violation by two units, and in turn that the neutrinos are Majorana particles \cite{Schechter:1981bd}. There is another cosmological observation which can distinguish a Dirac neutrino from a Majorana neutrino in a setup such as the left-right symmetric theory.  The anisotropy in the cosmic microwave background (CMB) is sensitive to extra radiation density arising from new light degrees of freedom that were in thermal equilibrium at some epoch with the standard model plasma.  The effect of such light particles is parametrized as $\Delta N_{\rm eff}$, and is measured in units of extra neutrino degrees of freedom.  The Planck experiment has measured $N_{\rm eff} = 2.99 \pm 0.17$ which includes baryon acoustic oscillation data.  In the near future, CMB-S4 \cite{Lee:2013bxd,Abazajian:2019eic} and Simon South Pole Observatory
\cite{SimonsObservatory:2019qwx} will explore $N_{\rm eff}$ with a sensitivity of 0.12, while the proposed  SPT-3G \cite{SPT-3G:2014dbx} will have a 3 sigma sensitivity of 0.06.  Such a precision could be used as a test of Dirac neutrino masses in contexts such as the left-right symmetric theories \cite{Abazajian:2019oqj}. 

A relativistic species that decouples from the plasma at a temperature $T_{\rm dec}$ would contribute to $\Delta N_{\rm eff}$ as
\begin{equation}
    \Delta N_{\mathrm{eff}} \simeq 0.027\left(\frac{106.75}{g_{\star}\left(T_{\mathrm{dec}}\right)}\right)^{4 / 3} g_{\rm eff}
    \label{eq:blue}
\end{equation}
where 
$g_{\rm eff}= (7/8) \times (2) \times (3) = 21/4$ accounts for the spin degree of the three $\nu_R$ states.  Now, as long as the reheat temperature after inflation is not too low compared to the $W_R$ gauge boson mass, the three $\nu_R$ fields would be in thermal equilibrium at a high enough temperature.  The decoupling temperature of the $\nu_R$'s can be estimated by equating their interaction rate with the Hubble expansion rate \cite{Nemevsek:2012cd}:
\begin{equation}
    G_F^2 \left(\frac{M_{W_L}}{M_{W_R}} \right)^4 T_{\rm dec}^5 \approx \sqrt{g^*(T_{\rm dec})}\, \frac{T_{\rm dec}^2}{M_{\rm Pl}},
\end{equation}
while yields
\begin{equation}
    T_{\rm dec} \simeq 400~ \mathrm{MeV}\left(\frac{g_{*}\left(T_{\rm dec}\right)}{70}\right)^{1 / 6}\left(\frac{M_{W_{R}}}{5~ \mathrm{TeV}}\right)^{4 / 3}~.
\end{equation}

We have plotted in Fig. \ref{fig:CMB} new contributions from the three $\nu_R$ fields to $\Delta N_{\rm efff}$ as a function of $T_{\rm dec}$. Also shown in the figure is the dependence on the $W_R$ gauge boson mass.  The solid blue line is a more exact form of Eq. (\ref{eq:blue}) obtained from the numerical results given in Ref. \cite{Abazajian:2019eic}. We see that with a sensitivity of 0.06 in $\Delta N_{\rm eff}$ the Dirac neutrino contributions can be measured or excluded, under the assumption that the reheat temperature after inflation is not orders of magnitude below the $W_R$ mass scale. 

It is worth noting that the idea of Dirac leptogenesis can be adopted in scenarios with Dirac neutrinos for explaining the baryon asymmetry of the universe \cite{Dick:1999je}. Here, even though there is no net lepton asymmetry induced, heavy particle decays or scatterings could  induce equal and opposite left-handed and right-handed lepton asymmetry. The electroweak sphalerons would convert part of the left-handed lepton asymmetry, but not the right-handed lepton asymmetry, to baryon asymmetry. This would lead to a net baryon asymmetry.  Of course, if neutrinos are Dirac particles, baryon number could still be violated in processes involving only quarks, which could also explain the observed baryon asymmetry of the universe.

\section{Realizing Pseudo-Dirac Neutrinos}
\label{sec:sec7}

In any context where the neutrino is naturally a Dirac fermion, there is a possibility that at a more fundamental level it is a pseudo-Dirac fermion. This could happen if quantum gravitational corrections, which are expected to conserve only gauge symmetries, generate tiny Majorana masses for the left-handed and/or the right-handed neutrinos.  In the model presented here, the leading operators of this type are
\begin{equation}
\frac{(\Psi_L \Psi_L \chi_L \chi_L)}{M_{\rm Pl}},~~ \frac{(\Psi_R \Psi_R \chi_R \chi_R)}{M_{\rm Pl}}~
\label{eq:Pl}
\end{equation}
which conserve all gauge quantum numbers and
 could be induced by quantum gravity.  If the coefficients of these operators are of order one, that would lead to a Majorana mass for the $\nu_R$ of order $10^{-2}$ eV (for $\kappa_R = 10$ TeV) and for $\nu_L$ of order $10^{-6}$ eV.  Such a value would result in a $\Delta m^2$ of order $10^{-3}~{\rm eV}^2$ for the squared mass splitting between active and sterile neutrinos.  Such values are excluded from cosmology \cite{Barbieri:1989ti,Enqvist:1990ek} as well as from solar neutrino data \cite{deGouvea:2009fp}, which require such mass splittings to be less than about $10^{-11}~{\rm eV}^2$. Here we propose a resolution to this potential problem.
 
 \begin{table}[!h]
    \centering
    \begin{tabular}{|c|c|c|c|c|c|c|c|c|c|c|}
    \hline
         & $Q_L$ & $Q_R$ & $\psi_L$ & $\psi_R$ & $P$  & $N$ & $E$ & $\chi_L$ & $\chi_R$ & $\varphi$ \\
         \hline
        $U(1)_X$ & 1/3 & 1/3 & $-1$ & $-1$ & $4/3$ & $-2/3$  & $-2$ & $1$ & $1$ & 0 \\
         \hline
         $U(1)_{B-L}$ & 1/3 & 1/3 & $-1$ & $-1$ & 1/3& 1/3  & $-1$ & 0 & 0  & $q$\\
         \hline
    \end{tabular}
    \caption{Charge assignment for fermions and scalars under $U(1)_X$ and $U(1)_{B-L}$.}
    \label{tab:BLcha}
\end{table}

As noted in Sec. \ref{sec:sec2}, the model presented has an anomaly-free $U(1)_{B-L}$ symmetry that may be gauged  without adding any new fermions.  Since quantum gravity corrections should respect this gauge symmetry, operators given in Eq.(\ref{eq:Pl}) will not be induced, as this would violate $(B-L)$. Note that under the $(B-L)$ symmetry, lepton fields have charge $-1$, while the Higgs doublets have charge zero.  The vector-like quarks have $(B-L)$ charge of $1/3$, while vector-like leptons have charge $-1$.  The full list of $(B-L)$ charges, along with the $U(1)_X$ charges of fermions and scalars are listed in Table \ref{tab:BLcha}. Note that with this charge assignment the Yukawa structure of Eq. (\ref{eq:chargedM}) remains intact even with the gauging of $(B-L)$ symmetry.  

If the $(B-L)$ symmetry is gauged and remains unbroken, the neutrinos will be strictly Dirac fermions. In this case, however, there are stringent limits on the $(B-L)$ gauge coupling arising from long-range forces.  It is preferable perhaps to break the symmetry spontaneously by the VEV of a singlet scalar $\varphi$. The $(B_L)$ charge of $\varphi$ is denoted as $q$, which is a rational number, $q \equiv m/n$, with $m,\,n$ being integers.  $\varphi$ carries no $U(1)_X$ charge, as indicated in Table  \ref{tab:BLcha}.

Once the scalar field $\varphi$ acquires a nonzero VEV, $U(1)_{B-L}$ would break down to a discrete $Z_{3m}$ subgroup.  This can be inferred by making all the charges under $(B-L)$ to be integers, which can be achieved by multiplying the charges listed in Table \ref{tab:BLcha} by $3n$, where the definition $q \equiv m/n$ is used.  With this normalization, the $(B-L)$ charge of $\varphi$ is $3m$, and consequently $Z_{3m}$ would remain unbroken once $\varphi$ acquires a VEV. While the operators of Eq. (\ref{eq:Pl}) would not be induced by quantum gravitational corrections, operators of the type
\begin{equation}
    \frac{(\Psi_L \Psi_L \chi_L \chi_L\varphi^k)}{M^{k+1}_{\rm Pl}},~~ \frac{(\Psi_R \Psi_R \chi_R \chi_R\varphi^k)}{M^{k+1}_{\rm Pl}}~
\label{eq:Pl1}
\end{equation}
could potentially be induced for some integer $k$.  However, this would require the following condition to be satisfied:
\begin{equation}
    -2+kq = 0,~~{\rm or}~~\frac{m}{n} = \frac{2}{k}~.
    \label{cod}
\end{equation}
Thus, if the charge of $\varphi$ is not of the form $q = 2/k$, no operator that violates lepton number by two units could be induced by Planck-scale corrections, and the neutrino would remain strictly a Dirac particle.

It is of interest to inquire possibilities when the $(B-L)$ charge of $\varphi$ is of the form $q=2/k$. For $k=1$, the operator $(\Psi_R \Psi_R \chi_R \chi_R \varphi/M^2_{\rm Pl})$ would be allowed. 
For other integer values of $k$, such as $k=2, \,3$, etc, the induced Majorana mass terms would be more suppressed.
The scale of $(B-L)$ symmetry breaking can be adjusted to obtain any desired mass-splitting between active and sterile neutrinos.  If $\Delta m^2$ is in the range of $(10^{-12} - 10^{-20})~{\rm eV}^2$, such pseudo-Dirac neutrinos may be tested in ultra high energy neutrinos at IceCube \cite{Keranen:2003xd,Beacom:2003eu} or in supernova neutrino signals \cite{Martinez-Soler:2021unz}.   The framework presented therefore provides a natural setup for pseudo-Dirac neutrinos as well.

\section{Conclusions}
\label{sec:sec8}

In this paper we have developed further a class of left-right symmetric models where the neutrinos are Dirac particles with naturally small masses \cite{Babu:1988yq}.  The gauge symmetry of these models is $SU(3)_c \times SU(2)_L \times SU(2)_R \times U(1)$ with the particles assigned to the gauge group in a left-right symmetric fashion.  This setup allows to identify parity as a good symmetry, which is only broken spontaneously and softly.  The charged fermion masses arise in this framework via their mixing with vector-like quarks and leptons, while the neutrino has no partner to mix with and thus remains massless at the tree-level. Quantum corrections induce small Dirac masses for the neutrinos.  We have evaluated the leading two-loop diagrams for the Dirac neutrino masses and shown the consistency of the framework with neutrino oscillation data. Both normal and inverted ordering of neutrino masses can be realized and the CP violating phase in neutrino oscillations is arbitrary.  

The structure of the quark mass matrices in this framework is such that it provides a solution to the strong CP problem based on parity symmetry, without the need for an axion.  The fermions of the model can arise naturally from $SU(5)_L \times SU(5)_R$ unification in its minimal version.  

We have shown how to preserve lepton number as a good symmetry in the neutrino mass matrix by gauging $(B-L)$ symmetry that is already anomaly free within the framework.  Upon spontaneous symmetry breaking, a discrete $Z_N$ subgroup of $(B-L)$ persists that prevents any $\Delta L = 2$ Planck-induced higher dimensional operators.  For certain values of $N$ of the discrete subgroup $Z_N$, the neutrino may be a pseudo-Dirac particle.  The amount of mass splitting between the active and sterile components of the neutrino is controlled by the scale of $(B-L)$ symmetry breaking.  We have also noted that these models can be tested in cosmological measurements, especially with high precision measurement of $\Delta N_{\rm eff}$, which is predicted to be $\Delta N_{\rm eff} \simeq 0.14$.

\vspace{-1mm}
\section*{Acknowledgements}
\quad We thank Julian Heeck for useful discussions. The work of KSB is supported in part by the US Department of Energy grant No. DE-SC 0016013. The work of XGH is supported in part by the MOST grant No. MOST 109-2112-M-002-017-MY3. KSB thanks KITP for the hospitality where part of this work was done. The research at KITP was
supported in part by the National Science Foundation under Grant No. NSF PHY-1748958. 
\bibliographystyle{utphys}
\bibliography{references}

\providecommand{\href}[2]{#2}\begingroup\raggedright\begin{thebibliography}{10}

\bibitem{Minkowski:1977sc}
P.~Minkowski, ``{$\mu \to e\gamma$ at a Rate of One Out of $10^{9}$ Muon
  Decays?},'' \href{http://dx.doi.org/10.1016/0370-2693(77)90435-X}{{\em Phys.
  Lett. B} {\bfseries 67} (1977) 421--428}.

\bibitem{Gell-Mann:1979vob}
M.~Gell-Mann, P.~Ramond, and R.~Slansky, ``{Complex Spinors and Unified
  Theories},'' {\em Conf. Proc. C} {\bfseries 790927} (1979) 315--321,
  \href{http://arxiv.org/abs/1306.4669}{{\ttfamily arXiv:1306.4669 [hep-th]}}.

\bibitem{Mohapatra:1979ia}
R.~N. Mohapatra and G.~Senjanovic, ``{Neutrino Mass and Spontaneous Parity
  Nonconservation},'' \href{http://dx.doi.org/10.1103/PhysRevLett.44.912}{{\em
  Phys. Rev. Lett.} {\bfseries 44} (1980) 912}.

\bibitem{Yanagida:1980xy}
T.~Yanagida, ``{Horizontal Symmetry and Masses of Neutrinos},''
  \href{http://dx.doi.org/10.1143/PTP.64.1103}{{\em Prog. Theor. Phys.}
  {\bfseries 64} (1980) 1103}.

\bibitem{Glashow:1979nm}
S.~L. Glashow, ``{The Future of Elementary Particle Physics},''
  \href{http://dx.doi.org/10.1007/978-1-4684-7197-7_15}{{\em NATO Sci. Ser. B}
  {\bfseries 61} (1980) 687}.

\bibitem{Dolinski:2019nrj}
M.~J. Dolinski, A.~W.~P. Poon, and W.~Rodejohann, ``{Neutrinoless Double-Beta
  Decay: Status and Prospects},''
  \href{http://dx.doi.org/10.1146/annurev-nucl-101918-023407}{{\em Ann. Rev.
  Nucl. Part. Sci.} {\bfseries 69} (2019) 219--251},
  \href{http://arxiv.org/abs/1902.04097}{{\ttfamily arXiv:1902.04097
  [nucl-ex]}}.

\bibitem{Babu:1988yq}
K.~S. Babu and X.~G. He, ``{Dirac neutrino masses as two loop radiative
  corrections},'' \href{http://dx.doi.org/10.1142/S0217732389000095}{{\em Mod.
  Phys. Lett. A} {\bfseries 4} (1989) 61}.

\bibitem{Pati:1974yy}
J.~C. Pati and A.~Salam, ``{Lepton Number as the Fourth Color},''
  \href{http://dx.doi.org/10.1103/PhysRevD.10.275}{{\em Phys. Rev. D}
  {\bfseries 10} (1974) 275--289}. [Erratum: Phys.Rev.D 11, 703--703 (1975)].

\bibitem{Mohapatra:1974gc}
R.~N. Mohapatra and J.~C. Pati, ``{A Natural Left-Right Symmetry},''
  \href{http://dx.doi.org/10.1103/PhysRevD.11.2558}{{\em Phys. Rev. D}
  {\bfseries 11} (1975) 2558}.

\bibitem{Mohapatra:1974hk}
R.~N. Mohapatra and J.~C. Pati, ``{Left-Right Gauge Symmetry and an
  Isoconjugate Model of CP Violation},''
  \href{http://dx.doi.org/10.1103/PhysRevD.11.566}{{\em Phys. Rev. D}
  {\bfseries 11} (1975) 566--571}.

\bibitem{Senjanovic:1975rk}
G.~Senjanovic and R.~N. Mohapatra, ``{Exact Left-Right Symmetry and Spontaneous
  Violation of Parity},''
  \href{http://dx.doi.org/10.1103/PhysRevD.12.1502}{{\em Phys. Rev. D}
  {\bfseries 12} (1975) 1502}.

\bibitem{Davidson:1987mh}
A.~Davidson and K.~C. Wali, ``{Universal Seesaw Mechanism?},''
  \href{http://dx.doi.org/10.1103/PhysRevLett.59.393}{{\em Phys. Rev. Lett.}
  {\bfseries 59} (1987) 393}.

\bibitem{Davidson:1987tr}
A.~Davidson and K.~C. Wali, ``{Family Mass Hierarchy From Universal Seesaw
  Mechanism},'' \href{http://dx.doi.org/10.1103/PhysRevLett.60.1813}{{\em Phys.
  Rev. Lett.} {\bfseries 60} (1988) 1813}.

\bibitem{Babu:1989rb}
K.~S. Babu and R.~N. Mohapatra, ``{A Solution to the Strong {CP} Problem
  Without an Axion},'' \href{http://dx.doi.org/10.1103/PhysRevD.41.1286}{{\em
  Phys. Rev. D} {\bfseries 41} (1990) 1286}.

\bibitem{Babu:2018vrl}
K.~S. Babu, B.~Dutta, and R.~N. Mohapatra, ``{A theory of R(D$^{*}$, D) anomaly
  with right-handed currents},''
  \href{http://dx.doi.org/10.1007/JHEP01(2019)168}{{\em JHEP} {\bfseries 01}
  (2019) 168}, \href{http://arxiv.org/abs/1811.04496}{{\ttfamily
  arXiv:1811.04496 [hep-ph]}}.

\bibitem{Hall:2018let}
L.~J. Hall and K.~Harigaya, ``{Implications of Higgs Discovery for the Strong
  CP Problem and Unification},''
  \href{http://dx.doi.org/10.1007/JHEP10(2018)130}{{\em JHEP} {\bfseries 10}
  (2018) 130}, \href{http://arxiv.org/abs/1803.08119}{{\ttfamily
  arXiv:1803.08119 [hep-ph]}}.

\bibitem{Craig:2020bnv}
N.~Craig, I.~Garcia~Garcia, G.~Koszegi, and A.~McCune, ``{P not PQ},''
  \href{http://dx.doi.org/10.1007/JHEP09(2021)130}{{\em JHEP} {\bfseries 09}
  (2021) 130}, \href{http://arxiv.org/abs/2012.13416}{{\ttfamily
  arXiv:2012.13416 [hep-ph]}}.

\bibitem{Mohapatra:1987hh}
R.~N. Mohapatra, ``{A Model for Dirac Neutrino Masses and Mixings},''
  \href{http://dx.doi.org/10.1016/0370-2693(87)90161-4}{{\em Phys. Lett. B}
  {\bfseries 198} (1987) 69--72}.

\bibitem{Lee:2016wiy}
C.-H. Lee and R.~N. Mohapatra, ``{Vector-Like Quarks and Leptons, SU(5)
  $\otimes$ SU(5) Grand Unification, and Proton Decay},''
  \href{http://dx.doi.org/10.1007/JHEP02(2017)080}{{\em JHEP} {\bfseries 02}
  (2017) 080}, \href{http://arxiv.org/abs/1611.05478}{{\ttfamily
  arXiv:1611.05478 [hep-ph]}}.

\bibitem{Schechter:1981bd}
J.~Schechter and J.~W.~F. Valle, ``{Neutrinoless Double beta Decay in SU(2) x
  U(1) Theories},'' \href{http://dx.doi.org/10.1103/PhysRevD.25.2951}{{\em
  Phys. Rev. D} {\bfseries 25} (1982) 2951}.

\bibitem{SPT-3G:2014dbx}
{\bfseries SPT-3G} Collaboration, B.~A. Benson {\em et~al.}, ``{SPT-3G: A
  Next-Generation Cosmic Microwave Background Polarization Experiment on the
  South Pole Telescope},'' \href{http://dx.doi.org/10.1117/12.2057305}{{\em
  Proc. SPIE Int. Soc. Opt. Eng.} {\bfseries 9153} (2014) 91531P},
  \href{http://arxiv.org/abs/1407.2973}{{\ttfamily arXiv:1407.2973
  [astro-ph.IM]}}.

\bibitem{SimonsObservatory:2019qwx}
{\bfseries Simons Observatory} Collaboration, M.~H. Abitbol {\em et~al.},
  ``{The Simons Observatory: Astro2020 Decadal Project Whitepaper},'' {\em
  Bull. Am. Astron. Soc.} {\bfseries 51} (2019) 147,
  \href{http://arxiv.org/abs/1907.08284}{{\ttfamily arXiv:1907.08284
  [astro-ph.IM]}}.

\bibitem{Lee:2013bxd}
{\bfseries Topical Conveners: K.N. Abazajian, J.E. Carlstrom, A.T. Lee}
  Collaboration, K.~N. Abazajian {\em et~al.}, ``{Neutrino Physics from the
  Cosmic Microwave Background and Large Scale Structure},''
  \href{http://dx.doi.org/10.1016/j.astropartphys.2014.05.014}{{\em Astropart.
  Phys.} {\bfseries 63} (2015) 66--80},
  \href{http://arxiv.org/abs/1309.5383}{{\ttfamily arXiv:1309.5383
  [astro-ph.CO]}}.

\bibitem{Abazajian:2019eic}
K.~Abazajian {\em et~al.}, ``{CMB-S4 Science Case, Reference Design, and
  Project Plan},'' \href{http://arxiv.org/abs/1907.04473}{{\ttfamily
  arXiv:1907.04473 [astro-ph.IM]}}.

\bibitem{Wolfenstein:1981kw}
L.~Wolfenstein, ``{Different Varieties of Massive Dirac Neutrinos},''
  \href{http://dx.doi.org/10.1016/0550-3213(81)90096-1}{{\em Nucl. Phys. B}
  {\bfseries 186} (1981) 147--152}.

\bibitem{Petcov:1982ya}
S.~T. Petcov, ``{On Pseudodirac Neutrinos, Neutrino Oscillations and
  Neutrinoless Double beta Decay},''
  \href{http://dx.doi.org/10.1016/0370-2693(82)91246-1}{{\em Phys. Lett. B}
  {\bfseries 110} (1982) 245--249}.

\bibitem{Valle:1983dk}
J.~W.~F. Valle and M.~Singer, ``{Lepton Number Violation With Quasi Dirac
  Neutrinos},'' \href{http://dx.doi.org/10.1103/PhysRevD.28.540}{{\em Phys.
  Rev. D} {\bfseries 28} (1983) 540}.

\bibitem{Kobayashi:2000md}
M.~Kobayashi and C.~S. Lim, ``{Pseudo Dirac scenario for neutrino
  oscillations},'' \href{http://dx.doi.org/10.1103/PhysRevD.64.013003}{{\em
  Phys. Rev. D} {\bfseries 64} (2001) 013003},
  \href{http://arxiv.org/abs/hep-ph/0012266}{{\ttfamily arXiv:hep-ph/0012266}}.

\bibitem{Beacom:2003eu}
J.~F. Beacom, N.~F. Bell, D.~Hooper, J.~G. Learned, S.~Pakvasa, and T.~J.
  Weiler, ``{PseudoDirac neutrinos: A Challenge for neutrino telescopes},''
  \href{http://dx.doi.org/10.1103/PhysRevLett.92.011101}{{\em Phys. Rev. Lett.}
  {\bfseries 92} (2004) 011101},
  \href{http://arxiv.org/abs/hep-ph/0307151}{{\ttfamily arXiv:hep-ph/0307151}}.

\bibitem{Keranen:2003xd}
P.~Keranen, J.~Maalampi, M.~Myyrylainen, and J.~Riittinen, ``{Effects of
  sterile neutrinos on the ultrahigh-energy cosmic neutrino flux},''
  \href{http://dx.doi.org/10.1016/j.physletb.2003.09.006}{{\em Phys. Lett. B}
  {\bfseries 574} (2003) 162--168},
  \href{http://arxiv.org/abs/hep-ph/0307041}{{\ttfamily arXiv:hep-ph/0307041}}.

\bibitem{Martinez-Soler:2021unz}
I.~Martinez-Soler, Y.~F. Perez-Gonzalez, and M.~Sen, ``{SN1987A still shining:
  A Quest for Pseudo-Dirac Neutrinos},''
  \href{http://arxiv.org/abs/2105.12736}{{\ttfamily arXiv:2105.12736
  [hep-ph]}}.

\bibitem{Lee:1956qn}
T.~D. Lee and C.-N. Yang, ``{Question of Parity Conservation in Weak
  Interactions},'' \href{http://dx.doi.org/10.1103/PhysRev.104.254}{{\em Phys.
  Rev.} {\bfseries 104} (1956) 254--258}.

\bibitem{Foot:1991py}
R.~Foot, H.~Lew, and R.~R. Volkas, ``{Possible consequences of parity
  conservation},'' \href{http://dx.doi.org/10.1142/S0217732392004031}{{\em Mod.
  Phys. Lett. A} {\bfseries 7} (1992) 2567--2574}.

\bibitem{Berezhiani:1995yi}
Z.~G. Berezhiani and R.~N. Mohapatra, ``{Reconciling present neutrino puzzles:
  Sterile neutrinos as mirror neutrinos},''
  \href{http://dx.doi.org/10.1103/PhysRevD.52.6607}{{\em Phys. Rev. D}
  {\bfseries 52} (1995) 6607--6611},
  \href{http://arxiv.org/abs/hep-ph/9505385}{{\ttfamily arXiv:hep-ph/9505385}}.

\bibitem{Silagadze:1995tr}
Z.~K. Silagadze, ``{Neutrino mass and the mirror universe},'' {\em Phys. Atom.
  Nucl.} {\bfseries 60} (1997) 272--275,
  \href{http://arxiv.org/abs/hep-ph/9503481}{{\ttfamily arXiv:hep-ph/9503481}}.

\bibitem{Farzan:2012sa}
Y.~Farzan and E.~Ma, ``{Dirac neutrino mass generation from dark matter},''
  \href{http://dx.doi.org/10.1103/PhysRevD.86.033007}{{\em Phys. Rev. D}
  {\bfseries 86} (2012) 033007},
  \href{http://arxiv.org/abs/1204.4890}{{\ttfamily arXiv:1204.4890 [hep-ph]}}.

\bibitem{Ma:2016mwh}
E.~Ma and O.~Popov, ``{Pathways to Naturally Small Dirac Neutrino Masses},''
  \href{http://dx.doi.org/10.1016/j.physletb.2016.11.027}{{\em Phys. Lett. B}
  {\bfseries 764} (2017) 142--144},
  \href{http://arxiv.org/abs/1609.02538}{{\ttfamily arXiv:1609.02538
  [hep-ph]}}.

\bibitem{Saad:2019bqf}
S.~Saad, ``{Simplest Radiative Dirac Neutrino Mass Models},''
  \href{http://dx.doi.org/10.1016/j.nuclphysb.2019.114636}{{\em Nucl. Phys. B}
  {\bfseries 943} (2019) 114636},
  \href{http://arxiv.org/abs/1902.07259}{{\ttfamily arXiv:1902.07259
  [hep-ph]}}.

\bibitem{Jana:2019mgj}
S.~Jana, P.~K. Vishnu, and S.~Saad, ``{Minimal realizations of Dirac neutrino
  mass from generic one-loop and two-loop topologies at $d = 5$},''
  \href{http://dx.doi.org/10.1088/1475-7516/2020/04/018}{{\em JCAP} {\bfseries
  04} (2020) 018}, \href{http://arxiv.org/abs/1910.09537}{{\ttfamily
  arXiv:1910.09537 [hep-ph]}}.

\bibitem{vanderBij:1983bw}
J.~van~der Bij and M.~J.~G. Veltman, ``{Two Loop Large Higgs Mass Correction to
  the rho Parameter},''
  \href{http://dx.doi.org/10.1016/0550-3213(84)90284-0}{{\em Nucl. Phys. B}
  {\bfseries 231} (1984) 205--234}.

\bibitem{Broadhurst:1987ei}
D.~J. Broadhurst, ``{The Master Two Loop Diagram With Masses},''
\href{http://dx.doi.org/10.1007/BF01551921}{{\em Z. Phys.} {\bfseries C47}
  (1990) 115--124}.

\bibitem{Ghinculov:1994sd}
A.~Ghinculov and J.~J. van~der Bij, ``{Massive two loop diagrams: The Higgs
  propagator},'' \href{http://dx.doi.org/10.1016/0550-3213(94)00522-G}{{\em
  Nucl. Phys. B} {\bfseries 436} (1995) 30--48},
  \href{http://arxiv.org/abs/hep-ph/9405418}{{\ttfamily arXiv:hep-ph/9405418}}.

\bibitem{Usyukina:1994eg}
N.~I. Usyukina and A.~I. Davydychev, ``{Two loop three point diagrams with
  irreducible numerators},''
  \href{http://dx.doi.org/10.1016/0370-2693(95)00136-9}{{\em Phys. Lett.}
  {\bfseries B348} (1995) 503--512},
\href{http://arxiv.org/abs/hep-ph/9412356}{{\ttfamily arXiv:hep-ph/9412356
  [hep-ph]}}.

\bibitem{McDonald:2003zj}
K.~L. McDonald and B.~H.~J. McKellar, ``{Evaluating the two loop diagram
  responsible for neutrino mass in Babu's model},''
  \href{http://arxiv.org/abs/hep-ph/0309270}{{\ttfamily arXiv:hep-ph/0309270}}.

\bibitem{Babu:2020bgz}
K.~S. Babu and A.~Thapa, ``{Left-Right Symmetric Model without Higgs
  Triplets},'' \href{http://arxiv.org/abs/2012.13420}{{\ttfamily
  arXiv:2012.13420 [hep-ph]}}.

\bibitem{tHooft:1972tcz}
G.~'t~Hooft and M.~J.~G. Veltman, ``{Regularization and Renormalization of
  Gauge Fields},''
\href{http://dx.doi.org/10.1016/0550-3213(72)90279-9}{{\em Nucl. Phys.}
  {\bfseries B44} (1972) 189--213}.

\bibitem{Coleman:1965xm}
S.~Coleman and R.~E. Norton, ``{Singularities in the physical region},''
\href{http://dx.doi.org/10.1007/BF02750472}{{\em Nuovo Cim.} {\bfseries 38}
  (1965) 438--442}.

\bibitem{Esteban:2020cvm}
I.~Esteban, M.~C. Gonzalez-Garcia, M.~Maltoni, T.~Schwetz, and A.~Zhou, ``{The
  fate of hints: updated global analysis of three-flavor neutrino
  oscillations},'' \href{http://dx.doi.org/10.1007/JHEP09(2020)178}{{\em JHEP}
  {\bfseries 09} (2020) 178}, \href{http://arxiv.org/abs/2007.14792}{{\ttfamily
  arXiv:2007.14792 [hep-ph]}}.

\bibitem{Planck:2018vyg}
{\bfseries Planck} Collaboration, N.~Aghanim {\em et~al.}, ``{Planck 2018
  results. VI. Cosmological parameters},''
  \href{http://dx.doi.org/10.1051/0004-6361/201833910}{{\em Astron. Astrophys.}
  {\bfseries 641} (2020) A6}, \href{http://arxiv.org/abs/1807.06209}{{\ttfamily
  arXiv:1807.06209 [astro-ph.CO]}}. [Erratum: Astron.Astrophys. 652, C4
  (2021)].

\bibitem{Abazajian:2019oqj}
K.~N. Abazajian and J.~Heeck, ``{Observing Dirac neutrinos in the cosmic
  microwave background},''
  \href{http://dx.doi.org/10.1103/PhysRevD.100.075027}{{\em Phys. Rev. D}
  {\bfseries 100} (2019) 075027},
  \href{http://arxiv.org/abs/1908.03286}{{\ttfamily arXiv:1908.03286
  [hep-ph]}}.

\bibitem{Nemevsek:2012cd}
M.~Nemevsek, G.~Senjanovic, and Y.~Zhang, ``{Warm Dark Matter in Low Scale
  Left-Right Theory},''
  \href{http://dx.doi.org/10.1088/1475-7516/2012/07/006}{{\em JCAP} {\bfseries
  07} (2012) 006}, \href{http://arxiv.org/abs/1205.0844}{{\ttfamily
  arXiv:1205.0844 [hep-ph]}}.

\bibitem{Dick:1999je}
K.~Dick, M.~Lindner, M.~Ratz, and D.~Wright, ``{Leptogenesis with Dirac
  neutrinos},'' \href{http://dx.doi.org/10.1103/PhysRevLett.84.4039}{{\em Phys.
  Rev. Lett.} {\bfseries 84} (2000) 4039--4042},
  \href{http://arxiv.org/abs/hep-ph/9907562}{{\ttfamily arXiv:hep-ph/9907562}}.

\bibitem{Barbieri:1989ti}
R.~Barbieri and A.~Dolgov, ``{Bounds on Sterile-neutrinos from
  Nucleosynthesis},''
  \href{http://dx.doi.org/10.1016/0370-2693(90)91203-N}{{\em Phys. Lett. B}
  {\bfseries 237} (1990) 440--445}.

\bibitem{Enqvist:1990ek}
K.~Enqvist, K.~Kainulainen, and J.~Maalampi, ``{Resonant neutrino transitions
  and nucleosynthesis},''
  \href{http://dx.doi.org/10.1016/0370-2693(90)91030-F}{{\em Phys. Lett. B}
  {\bfseries 249} (1990) 531--534}.

\bibitem{deGouvea:2009fp}
A.~de~Gouvea, W.-C. Huang, and J.~Jenkins, ``{Pseudo-Dirac Neutrinos in the New
  Standard Model},'' \href{http://dx.doi.org/10.1103/PhysRevD.80.073007}{{\em
  Phys. Rev. D} {\bfseries 80} (2009) 073007},
  \href{http://arxiv.org/abs/0906.1611}{{\ttfamily arXiv:0906.1611 [hep-ph]}}.

\end{thebibliography}\endgroup

\end{document}